\documentclass[11pt,a4paper]{article}
\usepackage{epsfig}

\newlength{\dinwidth}
\newlength{\dinmargin}
\def\fig#1{{Fig.~(\ref{#1})}}
\def\eq#1{{Eq.~(\ref{#1})}}
\newcommand{\Le}{\left(}
\newcommand{\Ra}{\right)}

\newcommand{\beq}{\begin{equation}}
\newcommand{\eeq}{\end{equation}}
\newcommand{\beqar}[1]{\begin{eqnarray}\label{#1}}
\newcommand{\eeqar}{\end{eqnarray}}
\newcommand{\ro}{\varrho}

\begin{document}

\title {{~}\\
{\Large \bf Regge Field Theory in zero transverse dimensions: \\
loops versus ''net'' diagrams}\\}
\author{ 
{~}\\
{~}\\
{\large 
%J.~Bartels$\,{}^{a)}\,$\thanks{Email: bartels@mail.desy.de},
%\hspace{1ex}
S.~Bondarenko$\,\,$\thanks{Email: sergeyb@ariel.ac.il}} 
\\[10mm]
{\it\normalsize  Ariel University Center, Israel}\\}

\maketitle
\thispagestyle{empty}

\begin{abstract}
Toy models of interacting Pomerons with  triple and quaternary Pomeron 
vertices in zero transverse dimension are investigated.
Numerical solutions for eigenvalues and eigenfunctions of the 
corresponding Hamiltonians are obtained, providing
the quantum solution for the 
scattering amplitude in both models.
The equations of motion for the  
Lagrangians of the theories are also considered  
and the classical solutions of the equations  are found.
Full two-point Green functions 
(''effective'' Pomeron propagator) and amplitude
of diffractive 
dissociation process are calculated in the framework of RFT-0 approach.
The importance of the loops contribution in the amplitude
at different values of the model parameters
is discussed as well as the difference between the models
with and without quaternary Pomeron vertex. 

\end{abstract}

%\begin{flushright}
%\vspace{-18.5cm}
%{\Large \bf DRAFT}\\
%\today
%\end{flushright}
%\thispagestyle{empty}

\newpage

\section{Introduction}

\,\,\,\,The complete understanding of the high energy scattering processes
is impossible without the calculation of 
the contributions of the different unitarization corrections
to the amplitude. In the QCD Pomeron framework, \cite{bfkl,bfklsum,nlbfkl}, 
these corrections are arose due the self Pomeron interactions via the
triple Pomeron interactions vertices \cite{vert1,vert2}. 
These self Pomeron interactions lead to the very complicated  picture of the 
amplitude's evolution with rapidity, see for example
\cite{eglla1,eglla2,eglla3,eglla4,jimwalk}.  

  There are few main approaches which claim that they properly describe the
Pomeron self interactions.
The first one is presented as a chain of the evolution equations
with Balitsky-Kovchegov (BK) equation as a main field equation of the hierarchy, \cite{BK}. 
The BK equation may be formulated in the framework of the Color Glass Condensate Approach
(CGK), \cite{jimwalk}, and may be properly analyzed in the terms
of s-channel interacting color dipoles, \cite{dipmod}. This equation  
correctly describes the interaction of two non identical
objects, for example  such interactions as DIS on nuclei, see 
\cite{bksols,bdepkov,jimsol,bksemi1,bksemi2,bk-pheno,kks,iim}.
Another approach is the QCD Reggeon Field Theory (RFT-QCD),
which is formulated in the momentum space and based on the
standard diagrammatic calculus, 
\cite{eglla1,eglla2,eglla3,eglla4,braun1,braun2,braun3}. In this approach the BK equation
describes a resummation of the ''fan'' diagrams of the type
depicted in Fig.1a with only Pomerons merging vertices considered. 

  Usual BK equation, describing "fan" diagrams, does not include  
Pomeron's splitting vertex. Therefore, the next natural 
step toward the unitarization of the amplitude is a  symmetrical consideration
of the scattering process with both vertices included. 
In this case the splitting vertex
may be accounted differently in two different approximations. In the first approach  
the semi-classical problem of the interest is considered and there the sum of the 
diagrams depicted in Fig.1b is calculated, neglecting Pomeron loops contribution into amplitude.
The second approach is concentrated on the solution of
the full quantum problem basing on some
effective high-energy QCD inspired models, accounting  
diagrams of Fig.1c type, see for example \cite{Lipatov:1995pn}.
So far several attempts of the calculations were done in both directions, on the basis of QCD-RFT
theory,  \cite{braun1,braun2,braun3} and in the framework of 
CGC and dipole model, \cite{selfdual}. The NLO and one-loop contribution into the amplitude also were calculated
, see for example \cite{NLO,OneLoop}, but the full solution still seems very far from it's completeness.
In this situation it is very
natural to consider much simpler zero 
transverse dimensional model, in hope that some important 
properties of solutions of this model will be faced in QCD as well.

  The RFT-0 (Reggeon Field Theory
in zero transverse dimensions) model is attracted much interest during few last years,
see   \cite{0dimsol}. The approach was formulated 
and studied a long time ago, even before the QCD era, see \cite{0dimc,0dimq,0dimi}.
The interest to the approach is due very attractive properties of the model.
First of all, RFT-0 is solvable at different parameters of the model.
This possibility to find full solution for the different regimes
of the theory is an important feature of the model, since we hope that we will understand
more about RFT-QCD if we will have full and semi-classical solutions for RFT-0,
see examples in \cite{BM,BondBr,Braun4}.
Therefore, the second important 
fact about the RFT-0 is that in this theory we can find both
classical and quantum solutions for the amplitude that provides us with the information
about the relative contribution of the loops to the amplitude.
The possibility to consider in RFT-0 
different types of the 
Pomeron vertices is an another property of the RFT-0 which is very useful and which could clarify the 
situation in QCD.

  In the paper we consider two  RFT-0 models, in the second one 
together with usual triple Pomeron vertex we also include
the quaternary Pomeron vertex. The paper is organized as follows. In the second section we 
consider the RFT-0 model with only triple Pomeron vertex.
In two subsections of the section we introduce an whole machinery of the problem for the both quantum
and classical solutions.
In the following subsections we present the results of our calculations
for the amplitude at different parameters of the model as well as the 
results for the two point Green's function (''effective'' Pomeron propagator )
of the theory. In the Sec.3 we solve the same problems
for the RFT-0  theory with the quaternary Pomeron vertex included.
The next section, Sec.4, is dedicated to the possible application
of the solution to the problem of diffractive dissociation.
The Sec.6 presents a summary of the results of the calculations and 
the last section, Sec.7, is the conclusion of our paper.

%%%%%%%%%%%%%%%%%%%%%%%%%%%%%%%

\section{Solution of RFT in zero transverse dimension with only triple Pomeron coupling}

 In this paper we investigate the one-dimensional problem of the interacting Pomerons
described by Lagrangian, which in the terms
of Gribov fields has the following form:
\beq\label{Lag1}
L\,=-\,\psi^{+}\,\dot{\psi}\,-\,\mu\,\psi^{+}\,\psi\,+\,i\,
\lambda\,\psi^{+}\,(\psi^{+}\,+\,\psi)\,\psi\,\,\,,
\eeq 
where $\,\mu\,$ is the intercept of the bare Gribov field,
$\lambda$ is the triple field interactions vertex
and differentiation means differentiation on rapidity (y)
variable, which is only the variable of the problem. 
Introducing the Pomeron fields, $q\,=\,i\,\psi^{+}\,$ and
$p\,=\,i\,\psi\,$ we rewrite the Lagrangian in the form
of real $\lambda\,$ coupling:
\beq\label{Lag2}
L\,=\,q\,\dot{p}\,+\,\mu\,q\,p\,-\,\lambda\,q\,(q\,+\,p)\,p\,\,.
\eeq 
The field q and p may be understood now in the terms
of the Pomeron creation and Pomeron annihilation operators $a^{+}$
$a$:
\beq\label{Oper}
\,q\,\rightarrow\,a^{+}\,=\,q\,\,\,\,,
\,p\,\rightarrow\,a\,=\,-\,\frac{d}{d\,q}\,.
\eeq
The Hamiltonian of the problem in the
terms of the operators q and p has the following form
\beq\label{Ham1}
H=-\,\mu\,q\,p\,+\,\lambda\,q\,(q\,+\,p)\,p\,\,\,,
\eeq
being
the second order differential operator:
\beq\label{Ham2}
H=\,(\mu\,q\,-\,\lambda\,q^{2})\,\frac{d}{dq}\,+
\,\lambda\,q\,\frac{d^2}{dq^2}\,\,.
\eeq
The full, quantum solution of the theory
will coincide, therefore, with the solution of the quantum mechanics problem
with the Hamiltonian given by Eq.\ref{Ham2}:
\beq\label{Ham3}
H\,\Psi\,=\,\frac{d\,\Psi}{d\,y}\,\,,
\eeq
where the function
\beq\label{Ham33}
\Psi(y,q)\,=\,\sum_{n=0}^{n=\infty}\,\lambda_{n}\,
e^{\,-\,E_{n}\,y}\,\phi_{n}(q)\,\,
\eeq
is the full quantum amplitude of the process of the scattering
of two particles. Here $E_{n}$ and $\phi_{n}(q)$
denote eigenvalues and eigenfunctions of the operator
Eq.~\ref{Ham2}, and  $\lambda_{n}$ are the normalized projections 
coefficients of the eigenfunctions on the value of $\,\Psi\,$ at $\,y=0\,$.

  The classical solution of the theory, i.e. the solution
with only ''tree'' diagrams without loops also may be 
obtained in the given framework. It is simply the solution of equations of motion 
for the Lagrangian given by Eq.~\ref{Lag2}

\begin{figure}[t]
\begin{center}
\psfig{file=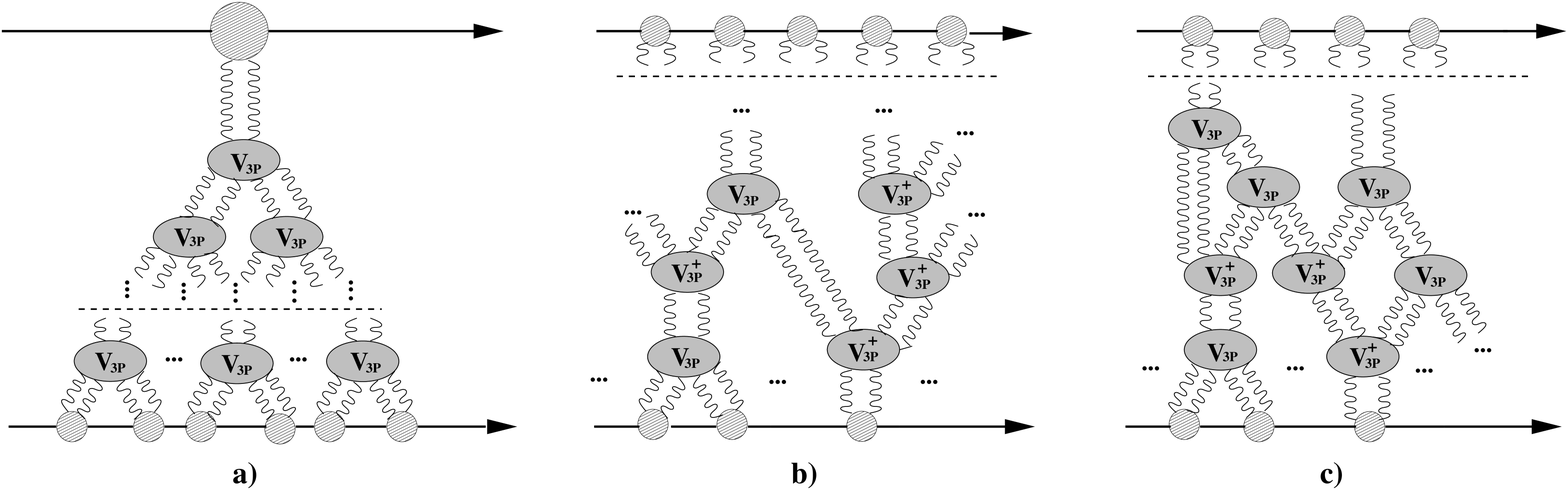,width=180mm} 
\end{center} 
\caption{\it 
Examples of diagrams of the effective theory of interacting Pomerons 
with triple pomeron vertices:
a) a fan diagram; b) a tree diagram defining the classical limit; 
c) a diagram with quantum loops.}
\label{diag1}
\end{figure}

%%%%%%%%%%%%%%%%%%%%%%%%%%%%%%%%%%%%%%%%%%%%%%%%%%%%%%%%%%%%%%%%%%%%%%%%%

\subsection{The quantum solution of the model}

 An approximate solution of the Eq.\ref{Ham3} was found many years ago,
see \cite{0dimc,0dimq,0dimi}. Analytical solutions of the Eq.\ref{Ham3} 
at large value of the ratio $\,\mu\,/\,\lambda\,\equiv\,\ro\,$
for ground state and first exited state were considered at the same time as well. Nevertheless,
the spectrum of the theory and eigenfunctions for next exited
states at large value of  $\,\ro\,$ as well as the solutions for 
arbitrary values of $\,\ro\,$ were not obtained. 
Therefore, we want to  define the procedure for the
numerical calculations of 
the spectrum of the theory and corresponding eigenfunctions
for the cases of the strong triple Pomeron coupling (small $\,\ro\,$  ) 
and week coupling (large $\,\ro\,$ ) as well. In the following
we will be more interesting in the case of the small triple coupling
as the case which is mostly correspond to the real perturbative QCD situation. 
So, we need to solve
the second order differential equation (see more detailed
in  \cite{0dimq}):
\beq\label{Quant1}
(\ro\,-\,q)\,\frac{d\,\Psi(y,q)}{dq}\,+
\,\frac{d^2\,\Psi(y,q)}{dq^2}\,=\frac{1}{\lambda\,q}\,
\frac{d\,\Psi(y,q)}{d\,y}\,\,,
\eeq
with the initial and boundary conditions on the function $\Psi(y,q)$:
\begin{eqnarray}\label{Quant2}
\,&\,&\,\Psi(y=0,q)\,=\,\sum_{n=0}^{n=\infty}\,\lambda_{n}\,
\,\phi_{n}(q)\,=\,I(q)\,\\
\,&\,&\,\Psi(y,q\,\rightarrow\,0)\,\propto\,q\,\\
\,&\,&\,\Psi(y,q\,\rightarrow\,\infty)\,\propto \,const.\,\,,
\end{eqnarray}
where form of $I(q)$ 
depends on the particularly considered physical problem. 
Now it is constructive to make a 
transformation of the eigenfunctions:
\beq\label{Quant3}
\phi_{n}(q)\,=\,e^{(q-\ro)^2\,/4}\,f_{n}(q)\,\,.
\eeq 
In this case the term which is proportional to the first derivative over q
is eliminated leading to hermitian Hamiltonian for the problem.
After this transformation we obtain a Shredinger type eigenvalue equation:
\beq\label{Quant4}
\frac{d^2\,f_{n}(q)}{dq^2}\,+\,\frac{f_{n}(q)}{2}\,-\,
\frac{1}{4}\,(q\,-\,\ro)^2\,f_{n}(q)\,=\,-\,\frac{E_n}{\lambda\,q}\,
f_{n}(q)\,
\eeq
with the following boundary conditions on the function $f_{n}(q)$:
\begin{eqnarray}\label{Quant5}
\,&\,&\,f_n(q\,\rightarrow\,0)\,\propto\,q\,\\
\,&\,&\,f_n(q\,\rightarrow\,\infty)\,\propto \,
e^{-q^2\,/\,4\,+\,q\ro\,/\,2}\,\,.
\end{eqnarray}
Our problem of interest is the scattering of two particles,
therefore we define initial condition for the $\Psi(y=0,q)$
in the eikonal form (see more in \cite{0dimq}) :
\beq\label{Quant6}
\Psi(y=0,q)\,=\,I(q)\,=\,I(q\,,q_{ext})\,=\,1-e^{-q\,q_{ext}}\,,
\eeq
here $q_{ext}$ is a value of the source
for the Pomeron field at zero rapidity. The
function $\Psi(y,q)$
results the evolution of interacting Pomeron fields for 
the value of the Pomeron-external particle vertex
equal to $q_{ext}$ at zero rapidity, till the 
value of the single Pomeron field equal to $q$ at rapidity y.

 The last ingredients of the theory are
projection coefficients $\,\lambda_n\,$ of the eigenfunctions
of the solution on the initial state I(q).
Having in mind, that our eigenfunctions are
orthogonal on interval of  q from $0$ to $\,\infty\,$
with weight function $\,F_{W}(q,\ro)\,$ :
\beq\label{Quant7}
\int_{0}^{\infty}\,f_{n}(q)\,f_{m}(q)\,F_{W}(q,\ro)\,dq\,=\,\delta_{n\,m}\,\,,
\eeq
we obtain for $\,\lambda_n\,$:
\beq\label{Quant8}
\lambda_n\,(q_{ext})=\,\frac{\int_{0}^{\infty}\,f_{n}(q)\,I(q\,,q_{ext})\,
F_{W}(q,\ro)\,dq\,
}{\int_{0}^{\infty}\,f_{n}^{2}(q)\,F_{W}(q,\ro)\,dq\,}\,\,,
\eeq
where the weight function $\,F_{W}(q,\ro)\,$ has the following form
(see \cite{0dimq} and also \cite{Heun}):
\beq\label{Quant9}
F_{W}(q,\ro)\,=\frac{\,e^{-(q\,-\,\ro)^2\,/\,2}}{\,q\,}\,.
\eeq
The numerical solution of Eq.\ref{Quant4}
with the boundary conditions given by Eq.\ref{Quant5}  
is the solution for the eigenvalues and eigenfunctions of the 
second order differential equation with two boundary value conditions.
These solutions for eigenfunctions and eigenvalues may be found
for each values of $\mu$ and $\lambda$, i.e. for the different
values of parameter $\,\ro\,$.
The obtained values of eigenfunctions $E_n$ and eigenfunctions $f_n(q)$
allow to complete the calculation of the full quantum scattering amplitude and , therefore, solve our problem. 
So, considering 
the scattering of two particles , where the vertex of interaction of the first
particle with Pomeron is equal $q_1$ and the vertex of interaction of 
the second particle with Pomeron is equal $q_2$, we
define the full , 
quantum solution for the scattering amplitude at given rapidity y as
\beq\label{Quant13}
\Psi(y,q=q_2)\,=\,\sum_{n=0}^{n=\infty}\,\lambda_{n}(q_1)\,
e^{\,-\,E_{n}\,y}\,e^{(q_2-\ro)^2\,/4}\,f_{n}(q_2)\,\,,
\eeq  
with the $\,\lambda_{n}(q_1)$ given by Eq.~\ref{Quant8}
for the known values of eigenfunctions, see Fig.1c.
The  solution for the amplitude, Eq.~\ref{Quant13},
does not depend from which value of Pomeron field, 
$q_1$ or $q_2$, the evolution begins.
If we consider the evolution of the Pomeron from $q_2$ to $q_1$ we will obtain 
the same answer for the amplitude, as for the case of 
evolution from $q_1$ to $q_2$.

%%%%%%%%%%%%%%%%%%%%%%%%%%%%%%%%%%%%%%%%%%%%%%%%%%%%%%%%%%%%%%%%%%%%%%%%%

\subsection{The classical solution of the model}

 In order to see the role 
of the Pomeron loops in the scattering amplitude
it is important to calculate the classical solution of our problem,
given by the ''net'' diagrams of Fig.1b.
The comparison between two solutions will show a relative weight
of the loops in the scattering amplitude and, therefore, an applicability of
the classical solution at given values of the theory parameters.
The required classical solution we obtain solving 
equations of motion for the Lagrangian given by Eq.\ref{Lag2} with the sources of 
Pomeron fields at zero rapidity and final rapidity of
the process Y:  
\beq\label{Class1}
L\,=\,\frac{1}{2}\,q\,\dot{p}\,-\frac{1}{2}\,\dot{q}\,p\,+
\,\mu\,q\,p\,-\,\lambda\,q\,(q\,+\,p)\,p\,+\,
\,q(y)\,p_0(y)\,+\,q_0(y)\,p(y)\,\,,
\eeq
where $\,p_0(y)\,\,q_0(y)\,$ are the sources of the Pomeron fields.
The equations of the motion for the fields p and q 
are the following:
\begin{eqnarray}\label{Class2}
\,&\,&\,\dot{q}\,=\,\mu\,q\,-\,\lambda\,q^2\,-\,2\,\lambda\,q\,p\,\\
\,&\,&\,\dot{p}\,=\,-\,\mu\,p\,+\,\lambda\,p^2\,+\,2\,\lambda\,q\,p\,\\
\,&\,&\,q_0(y)\,=\,q_1\,\delta(y)\,\\
\,&\,&\,p_0(y)\,=\,q_2\,\delta(y-Y)\,\,.
\end{eqnarray}
This system of equations is another example of two value boundary problem
for the system of first order differential equations, and may be also solved
for different values of parameters
$\mu,\,\,\lambda,\,\,q_1,\,\,q_2\,$. The solutions of the system
Eq.\ref{Class2} at given value of final rapidity Y and given values of
parameters $\mu,\,\,\lambda,\,\,q_1\,\,q_2\,$  we will
denote by $\{q_{c}(y),p_{c}(y)\}$. 
With the  solution  $\{q_{c}(y),p_{c}(y)\}$ the 
''net'', classical  amplitude , represented by
diagrams Fig.1b, is defined by standard way as:
\beq\label{Class3}
\Psi_{c}(Y,\,q_{c},\,p_{c})\,=\,
1\,-\,e^{-S(Y,\,q_{c},\,p_{c})}\,\,,
\eeq
where
\beq\label{Class4}
S(Y,q_{c},p_{c})=\int_{0}^{Y} L(Y,q_{c}(y),p_{c}(y)) dy = 
\frac{1}{2}(q_1 p_{c}(0)+q_{c}(Y)q_2)+
\frac{\lambda}{2}\int_{0}^{Y} (q^2_{c}(y)p_{c}(y)+
q_{c}(y)\,p^2_{c}(y)) dy\,.
\eeq
The amplitude Eq.\ref{Class3} describes
the eikonalized interactions of ''net'' diagrams
and for symmetric boundary conditions, $q_1=q_2$,
the amplitude is invariant under the duality transformations:
\beq\label{Class5}
p\,\leftrightarrow\,q\,\,\,\,and\,\,\,\,y\leftrightarrow\,Y-y\,.
\eeq

 The interesting feature of the classical solution of the system
defined by the
Lagrangian Eq.\ref{Class1} is that starting from some critical rapidity
$Y_c$ the solution of equation of motion is not unique
for $q_1=q_2\,<\,\ro$.
From the  rapidity $Y_c$ there are at least three  
classical trajectories  $\{q^{i}_{c}(y),p^{i}_{c}(y)\}$ which locally minimize the action and
with increasing of the rapidities the number of the trajectories is 
growing. Each from them provides the local minimum of the action and the 
amplitude of the theory, therefore, 
must be rewritten in the following form:
\beq\label{Class6}
\Psi_{c}(Y)\,=\,1\,-\,
\sum_{i}\,\Delta_{i}\,\exp\{-S(Y,\,q^{i}_{c}(y)\,,p^{i}_{c}(y))\}\,,
\eeq
where $\,\Delta_{i}\,$ is the quantum weight of the corresponding 
classical trajectories. In the following consideration
these weights we  take  equal 1 or -1, depending on the type 
of trajectory, see more detailed consideration of this problem in \cite{0dimi}.

\begin{figure}[t]
\begin{center}
\psfig{file=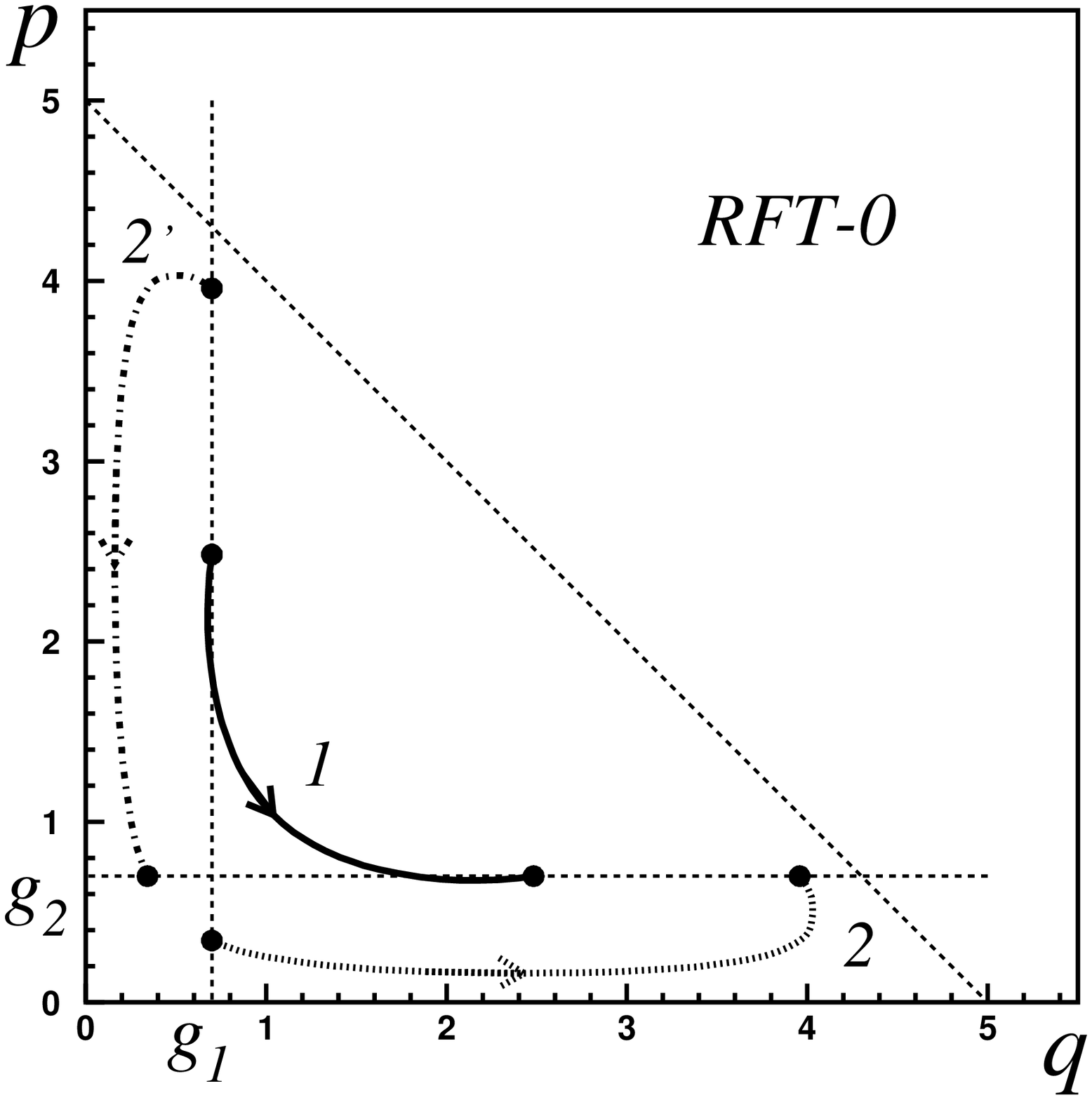,width=100mm} 
\end{center} 
\caption{\it Classical solutions of the RFT-0 with
only triple Pomeron vertex: 
 the $\{ q,p\}$ trajectories obtained for $Y=5\,>\,Y_c\,$, $q=p=0.7$, 
$\rho\,=\,5$.}
\label{Traj}
\end{figure}

  In our calculations of the classical amplitude for the symmetrical case
$q_1=q_2$,
we always will use three solutions of the equations of motion.
The first one, $\{q^{1}_{c}(y),p^{1}_{c}(y)\}$,  is the  symmetrical in the sence of Eq.\ref{Class5}, 
which trajectory will be similar to the trajectory number 1 in the  
Fig.\ref{Traj}. 
Another two dominant solutions, which we use in the calculations and 
which arise from some rapidity  $Y_c$ ,
are the solutions for the trajectories similar to the trajectories 
2 and 3 in the Fig.\ref{Traj}.
Separately each of them is not symmetrical in the sense of 
the symmetry transformation
given by the Eq.\ref{Class5}, but instead, under
the duality transformation of rapidity given by the Eq.\ref{Class5}, there is
a pair symmetry of these two solutions 
\beq
q^{2}_{c}(y)\,=\,p^{3}_{c}(Y-y)\,\,\,\,and\,\,\,\,p^{2}_{c}(y)\,=\,q^{3}_{c}(Y-y). 
\eeq
It is interesting to note, that asymptotic behavior of these 
two classical solutions is well known, each of them may be described by
''fan'' diagram amplitude of Fig.1a, see  \cite{0dimq},
and being the dominant contributions these solutions lead to the 
''fan'' dominance effect, see ~\cite{BM}.

  The  final solution of our problem in the classical approximation
we can write with the help of these three particular solutions as  
\beq\label{Class7}
\Psi_{c}(Y)=1+exp\{-S(Y,\,q^{1}_{c}(y)\,,p^{1}_{c}(y))\}-
exp\{-S(Y,\,q^{2}_{c}(y)\,,p^{2}_{c}(y))\}-
exp\{-S(Y,\,q^{3}_{c}(y)\,,p^{3}_{c}(y))\}\,,
\eeq
which is symmetrical in respect to the duality transformation Eq.\ref{Class5}.
In our plots for the symmetrical interactions, $q_1=q_2$,
the classical solution will be always presented 
by amplitude of Eq.\ref{Class7}, which contains three parts beginning from
critical rapidity  $Y_c$, and only one part, symmetrical solution, at
rapidities smaller than  $Y_c$.  In the asymmetrical case the symmetry is
broken initially, and, therefore, only one ''fan'' configuration survives
in the classical solution.

 It is important to underline, that
this picture for classical solution holds for each from the considered models,
for RFT-0 with triple Pomeron vertex only and for RFT-0 model with both
triple and quaternary vertexes.

%%%%%%%%%%%%%%%%%%%%%%%%%%%%%%%%%%%%%%

\subsection{Parameters of the model}

 First of all, we define the range and value of parameters
for which the calculations were done in both cases, quantum and classical.
We calculated quantum and classical 
amplitudes,
$\,\Psi(Y,\,q)\,$ and
$\,\Psi_{c}(Y,\,q_{c},\,p_{c})\,$, for the following value of 
$\ro\,=\,\mu\,/\,\lambda\,$:
\begin{eqnarray}\label{Num1}
%\,&\,&\,\ro\,=\,1\,\,\,\,(\mu\,=\,0.1\,,\,\lambda\,=\,0.1\,);\\
%\,&\,&\,\ro\,=\,3\,\,\,\,(\mu\,=\,0.21\,,\,\lambda\,=\,0.07\,);\\
\,&\,&\,\ro\,=\,5\,\,\,\,(\mu\,=\,0.2\,,\,\lambda\,=\,0.04\,).
\end{eqnarray}
The reason of the main attention on this value of $\ro\,$ is very simple.
The coupling constant in RFT-0 may be defined as $\,\alpha_s\,=\,1\,/\,\ro$  and
at $\ro\,=\,5\,$ we have, therefore, $\,\alpha_s\,=\,0.2\,$ .
This  value for the coupling constant is reasonable also in QCD that provides the possibility of some analogies between two theories.
The value of the external sources we take as follow
\begin{itemize}\label{Num2}
\item symmetrical case:
\beq
\,q_1\,=\,q_2\,=0.1\,-\,1\,\,;
\eeq
\item non-symmetrical case:
\beq
\,q_1\,=\,0.2\,;\,\,\,q_2\,=0.3\,-\,1\,.
\eeq
\end{itemize}
As an example of the case of the strong coupling , i.e. large value of the triple vertex,
we also will present the quantum solution for the symmetrical case of
interaction at
\begin{eqnarray}\label{Num11}
\,&\,&\,\ro\,=\,1\,\,\,\,(\mu\,=\,0.1\,,\,\lambda\,=\,0.1\,).
\end{eqnarray}

\begin{figure}[hptb]
\begin{center}
\psfig{file=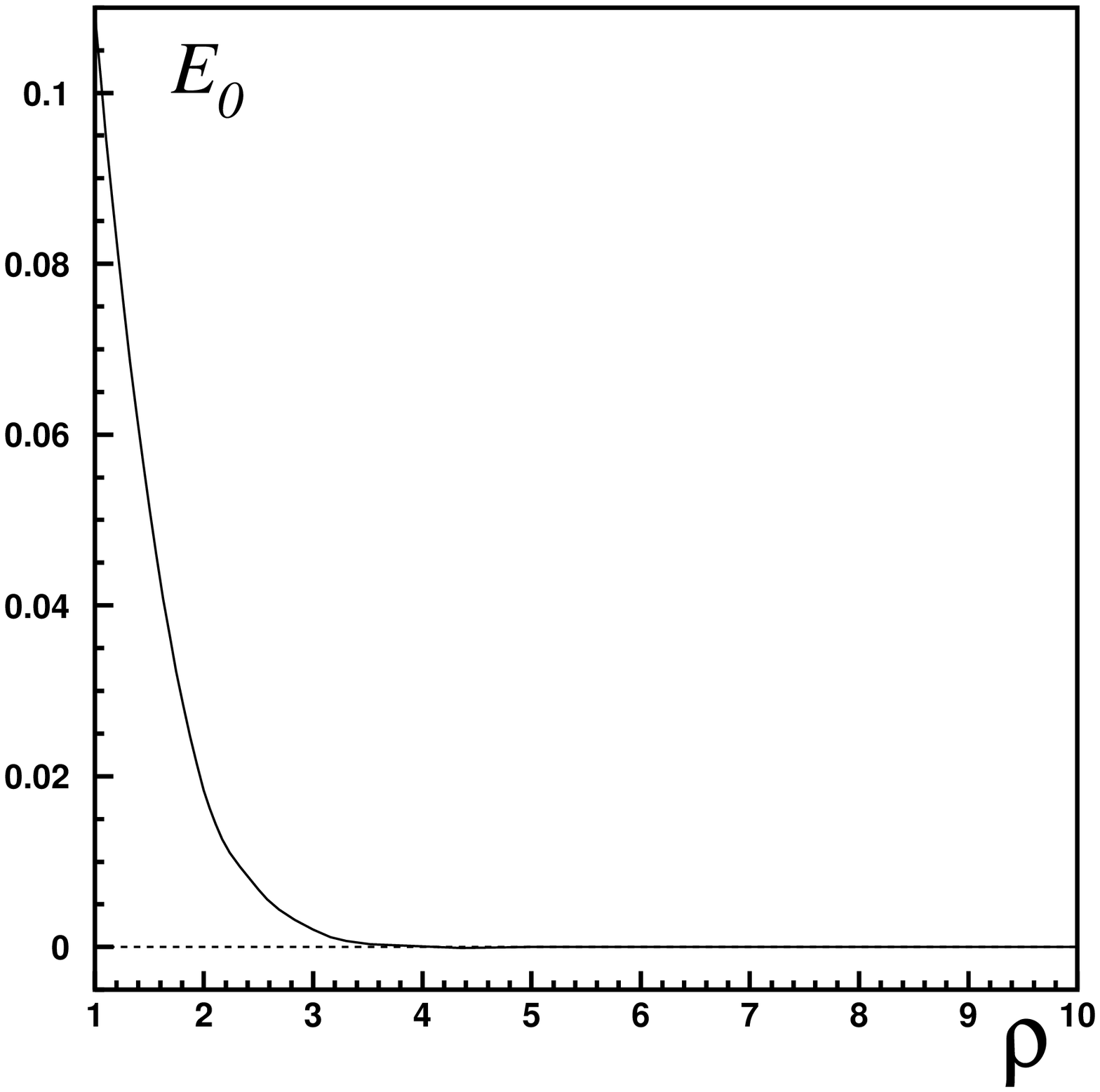,width=100mm}
\end{center} 
\caption{\it\large The ground state of RFT as a function of $\ro$.}
\label{EFUN77}
\end{figure}

 Now it is instructive to consider the spectrum of the RFT-0 theory
at different $\ro$. The Table\ref{EFUN} presents the found eigenvalues of the
theory, for $\,\ro\,=\,5\,$ as well as for $\,\ro\,=\,1\,$ and  for
$\,\ro\,=\,3\,$.
\begin{table}[t]
\begin{center}
\begin{tabular}{|c|c|c|c|c|c|c|c|c|c|c|}
\hline
\, & \,& \,& \, &\, &\, &\, &\, &\, &\, &\,\\
$\ro\,$ &  $E_0$ & $E_1$ & $E_2$ & $E_3$ &
 $E_4$ & $E_5$ & $E_6$ & $E_7$ & $E_8$ &
 $E_9$ \\
\, & \,& \,& \, &\, &\, &\, &\, &\, &\, & \\
\hline
\, & \,& \,& \, &\, &\, &\, &\, &\, &\, &\,\\
$\ro\,=\,1\,$ & 0.0546 & 0.292 & 0.609 & 0.997 &
 1.447 & 1.95 & 2.503 & 3.103 & 3.745 & 4.427  \\
\, & \,& \,& \, &\, &\, &\, &\, &\, &\, &\,\\
\hline
\, & \,& \,& \, &\, &\, &\, &\, &\, &\, &\, \\
$\ro\,=\,3\,$ & 0.00217 & 0.142 & 0.297 &
0.507 & 0.759 & 1.047 & 1.37 & 1.724 & 2.108 & 2.579\\
\, & \,& \,& \, &\, &\, &\, &\, &\, &\, &\,\\
\hline
\, & \,& \,& \, &\, &\, &\, &\, &\, &\, &\, \\
$\ro\,=\,5\,$ &  $ 1.36\cdot\,10^{-6}$ &
 0.144 & 0.178 & 0.276 &
 0.386 & 0.519 & 0.671 & 0.841 & 1.03 & 1.228 \\
\, & \,& \,& \, &\, &\, &\, &\, &\, &\, &\,\\
\hline
\end{tabular}
\caption{\it The eigenvalues of Eq.\ref{Quant4} for different values
of parameter $\ro$.}
\label{EFUN}
\end{center}
\end{table}
The asymptotic behavior 
of the amplitude is clearly seen from the Fig.\ref{EFUN77} which represents
the behavior of the ground state as a function of  $\,\ro\,$.
Being important at small rapidity, all exited states, i.e.
states with eigenvalues $E_i,\,\,\,i=1..\infty\,$, at larger rapidities
are decreasing very rapidly to zero. Therefore,
at asymptotically large rapidity only the ground state survives,
which nonzero origin is due the tunneling effect of
the order $e^{-\ro^2\,/\,2}$ which arises
between the Coulomb and harmonic wells of the potential
of the Eq.\ref{Quant4}, see \cite{0dimc,0dimq,0dimi}.
At large enough rapidity the  amplitude for any value of $\ro$ is zero:
for small $\ro$ it happens rapidly already at small rapidities, at large $\ro$ huge rapidity needs.
As we will see further, this is a particular property of the model
with only triple Pomeron vertex, the theory with additional quaternary vertex
has precisely zero ground state.

%%%%%%%%%%%%%%%%%%%%%%%%%%%%%%%%%%%%%%

\subsection{Numerical results: loops versus ''net'' diagrams}

\begin{figure}[hptb]
\begin{tabular}{ c c}
\psfig{file=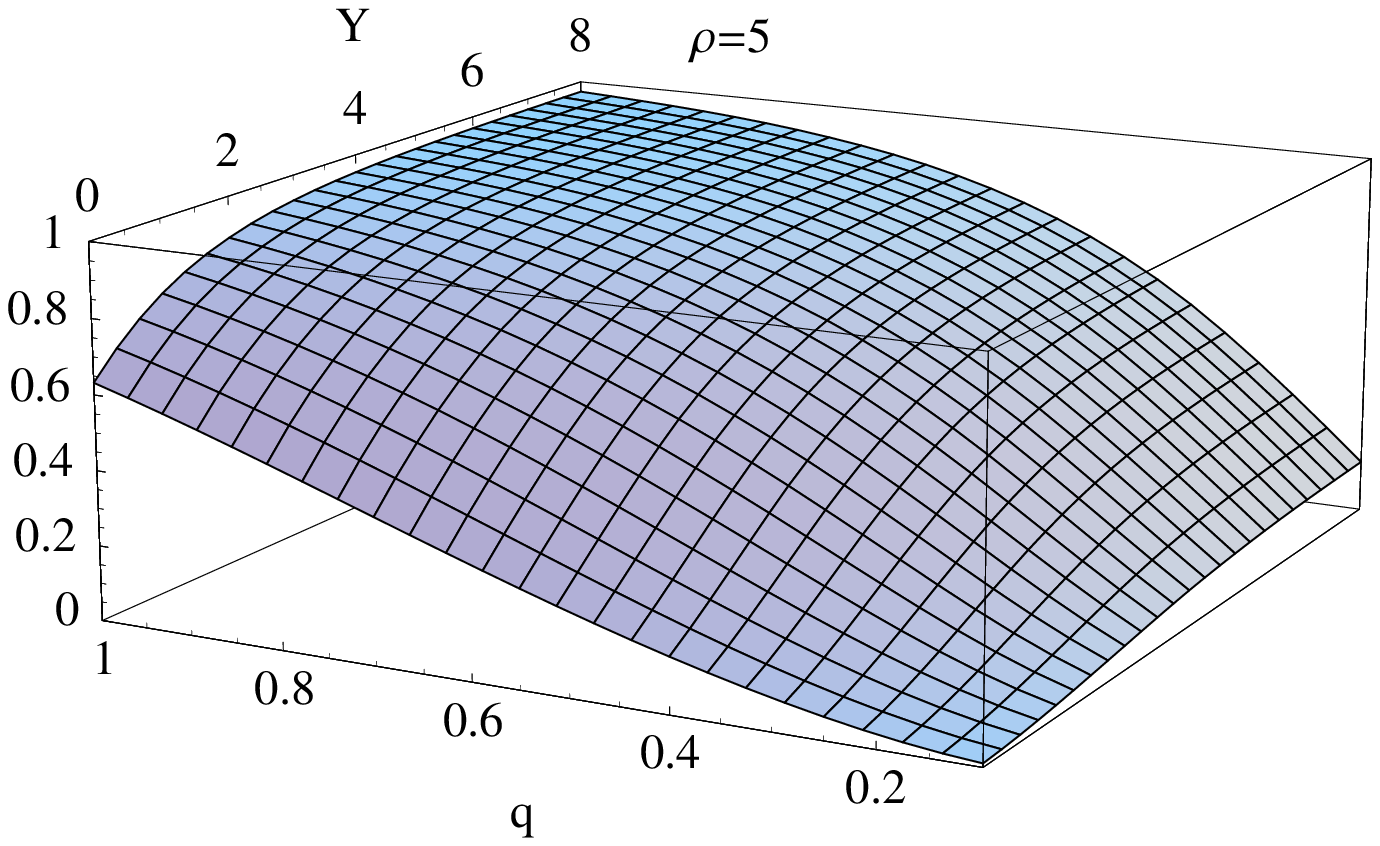,width=115mm} & 
\psfig{file=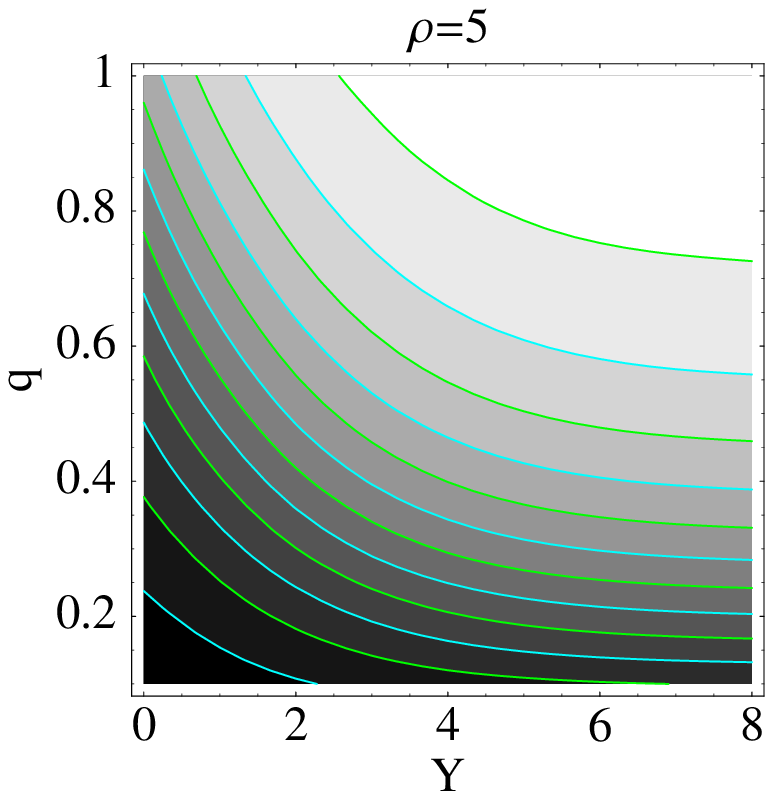 ,width=60mm}\\
 &  \\ 
\fig{EFUN3}-a & \fig{EFUN3}-b \\
 &  \\
\end{tabular}
\caption{\it The 
quantum amplitude  $\Psi(Y,\,q)\,$ of the 3P vertex model
in the form of 3D and contour plots
at $\ro\,=\,5\,$  as a functions of scaled
rapidity Y and symmetrical values of external sources $q_1=q_2$. }
\label{EFUN3}
\end{figure}

\begin{figure}[hptb]
\begin{tabular}{ c c}
\psfig{file=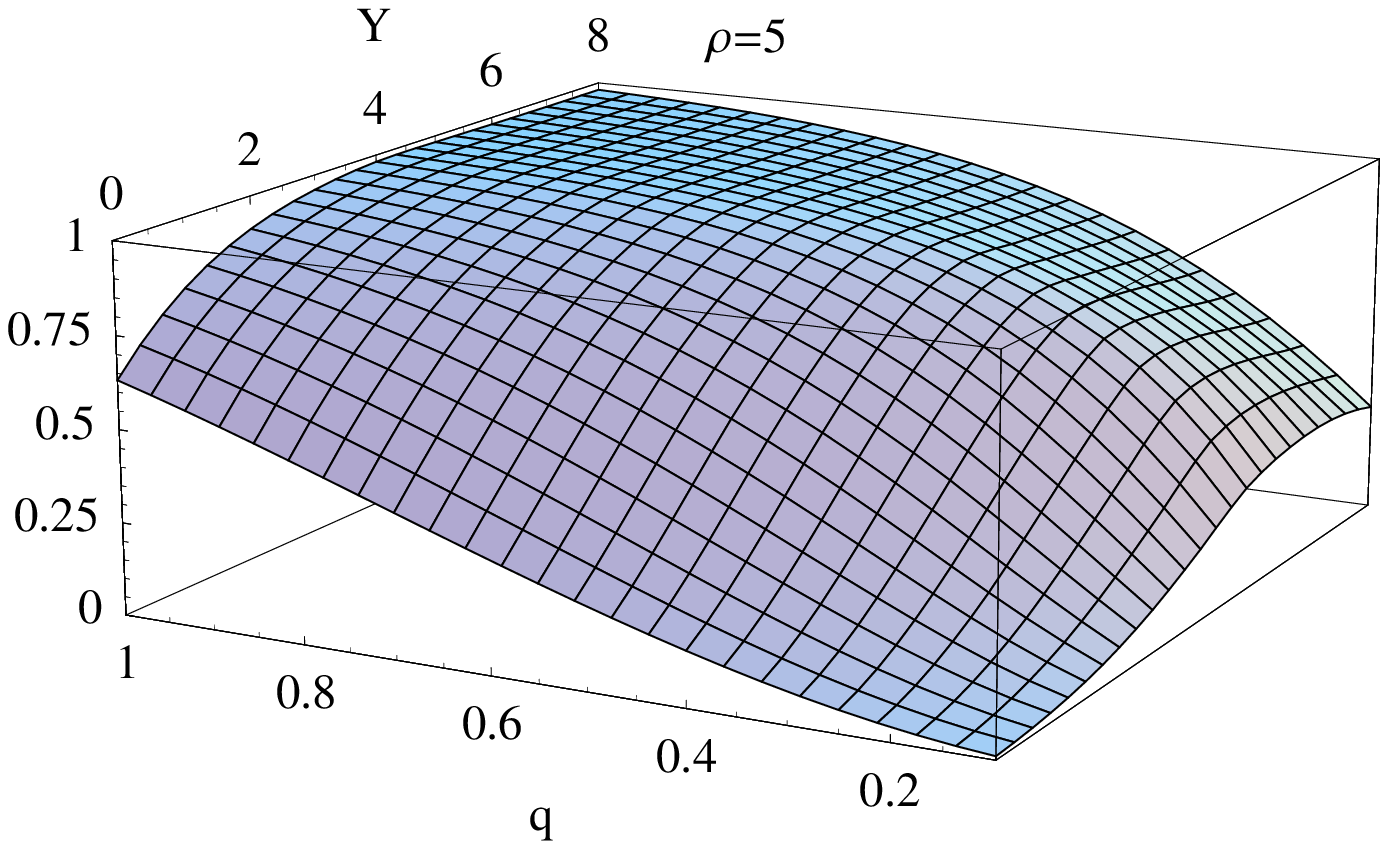,width=115mm} &
\psfig{file=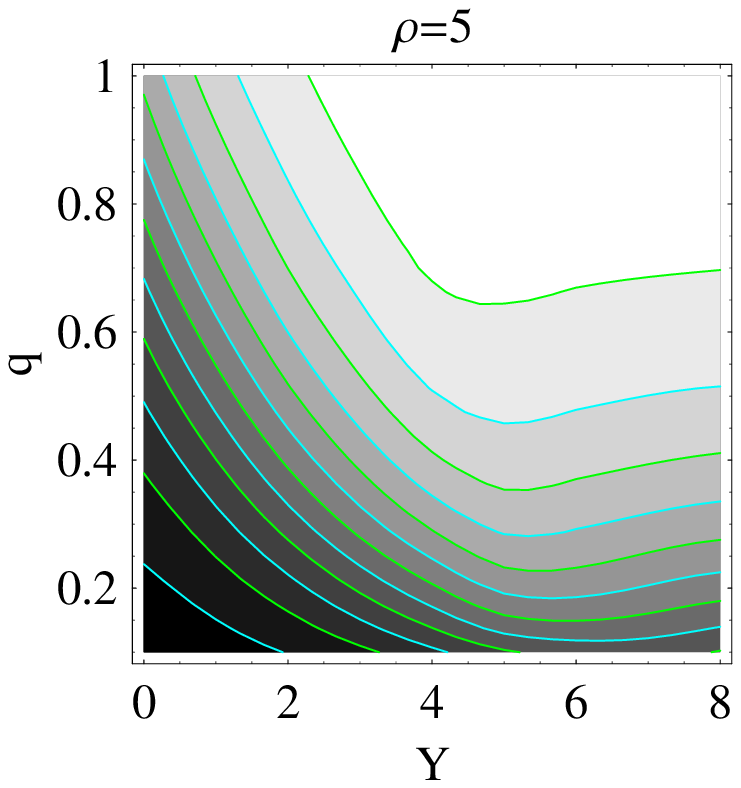,width=60mm}\\
&  \\ 
\fig{EFUN4}-a & \fig{EFUN4}-b \\
 &  \\
\end{tabular}
\caption{\it The 
classical amplitude  $\,\Psi_{c}(Y,\,q_{c},\,p_{c})\,$ 
of the 3P vertex model in the form of 3D and contour plots at
$\ro\,=\,5\,$ and presented as a functions of scaled
rapidity Y and symmetrical values of external sources $q_1=q_2$.}
\label{EFUN4}
\end{figure}

\begin{figure}[hptb]
\begin{tabular}{ c c}
\psfig{file=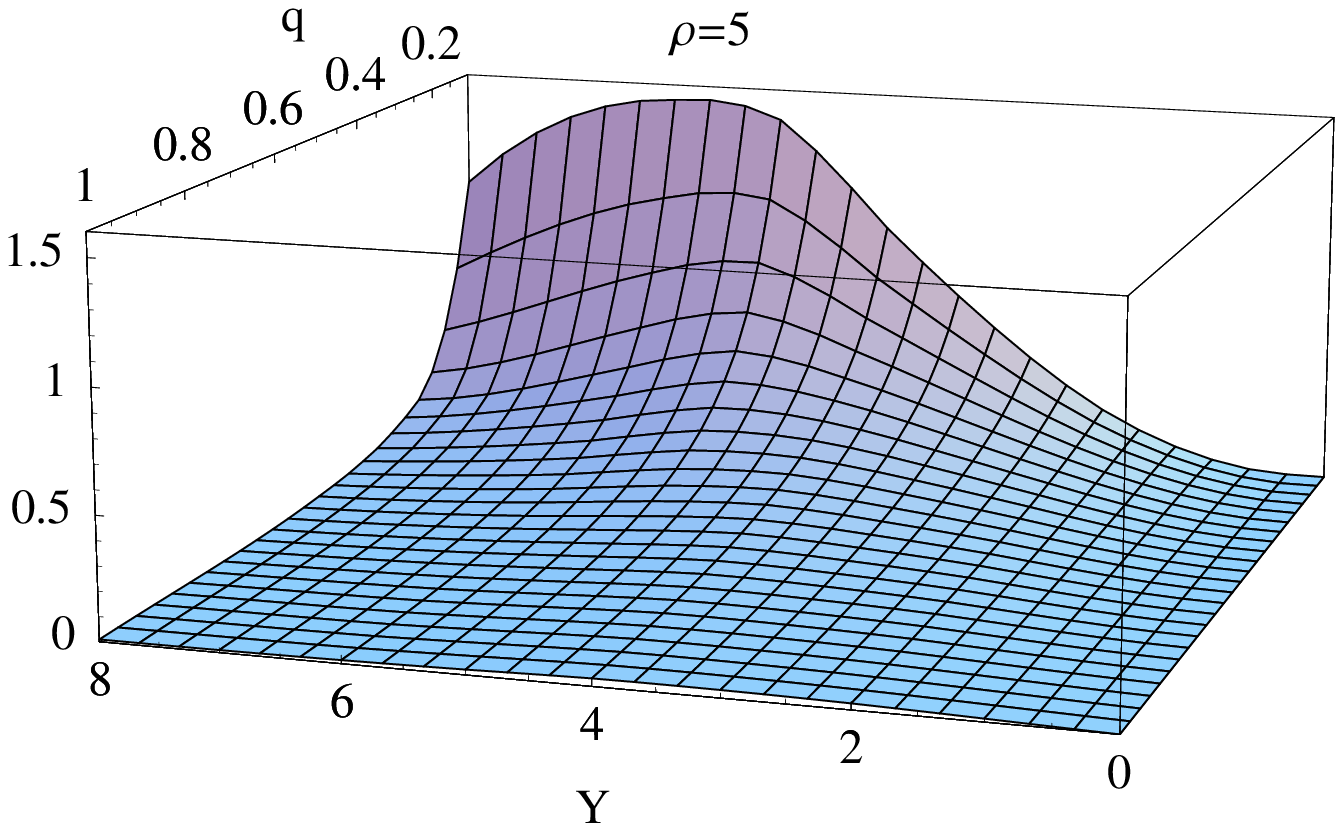,width=115mm} & 
\psfig{file=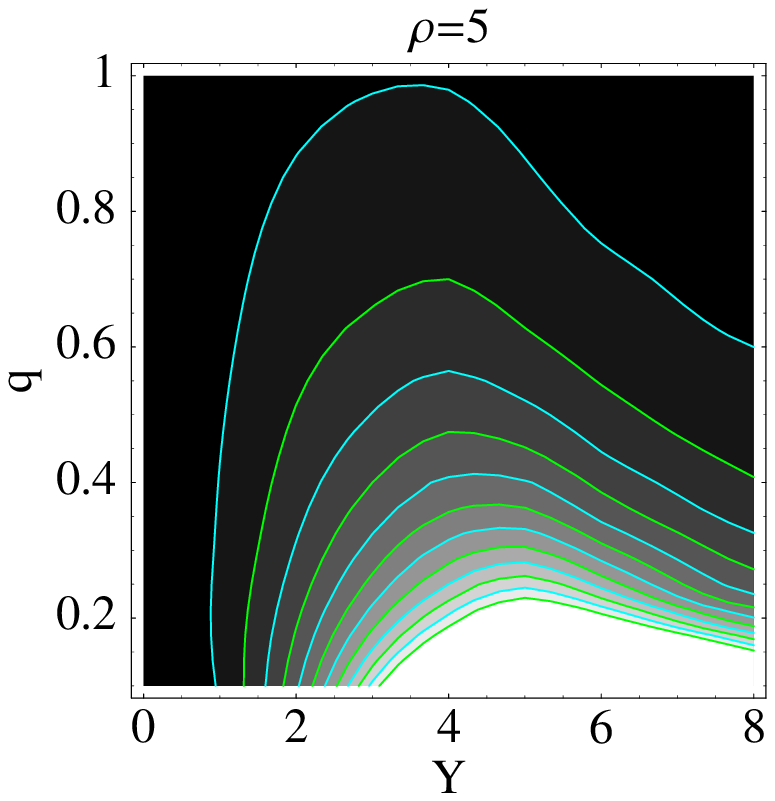 ,width=60mm}\\
 &  \\ 
\fig{EFUN5}-a & \fig{EFUN5}-b \\
 &  \\
\end{tabular}
\caption{\it
The ratio $|\Psi(Y,\,q)\,-\,\Psi_{c}(Y,\,q_{c},\,p_{c})|\,/\,\Psi(Y,\,q)$
in the 3P vertex model
is represented by 3D plot and contour plot. The plots are presented as a functions
of scaled rapidity Y at $\ro\,=\,5\,$ and at $q_1=q_2=0.2-1$.}
\label{EFUN5}
\end{figure}

\begin{figure}[hptb]
\begin{tabular}{ c c}
\psfig{file=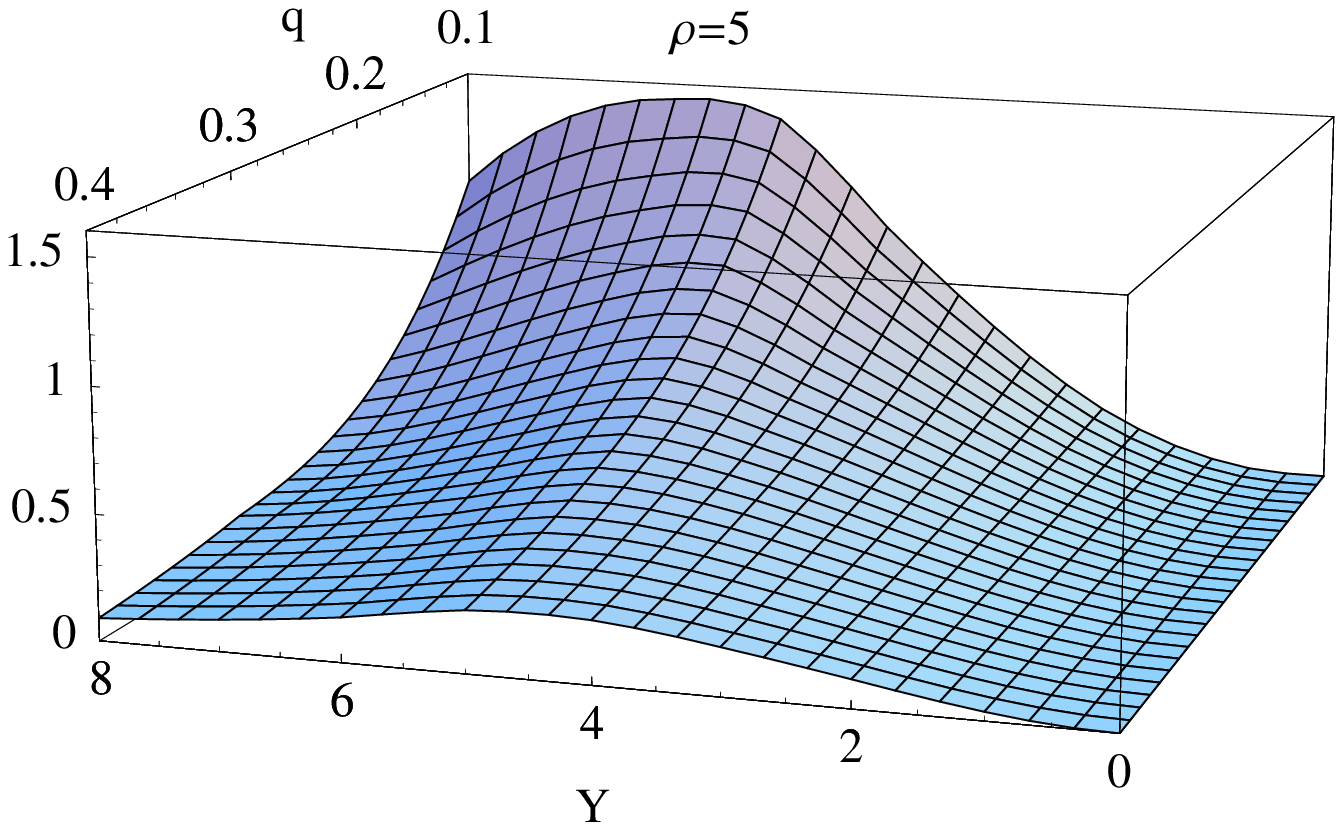,width=90mm} & 
\psfig{file=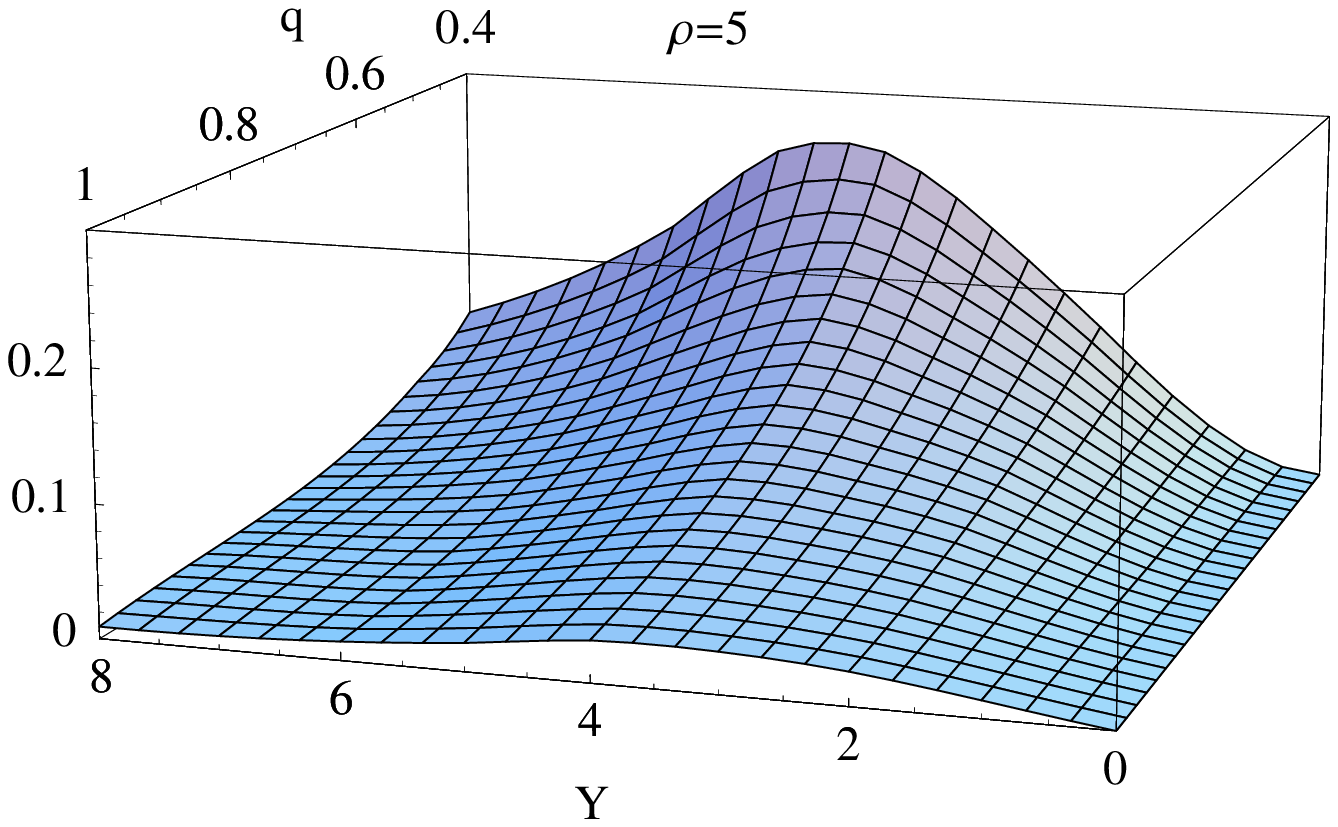,width=90mm} \\
 &  \\ 
\fig{EFUN6}-a & \fig{EFUN6}-b \\
 &  \\
\end{tabular}
\caption{\it 
The same ratio $|\Psi(Y,\,q)\,-\,\Psi_{c}(Y,\,q_{c},\,p_{c})|\,/\,\Psi(Y,\,q)$
in the 3P vertex model
is represented by 3D plots, now
at  the values of the sources $q_1=q_2=0.1-0.4$ in the Fig.7-a
and at  the values of the sources $q_1=q_2=0.4-1$ in the Fig.7-b
correspondingly. The plots are presented as a functions
of scaled rapidity Y at $\ro\,=\,5\,$ . }
\label{EFUN6}
\end{figure}

\begin{figure}[hptb]
\begin{tabular}{ c c}
\psfig{file=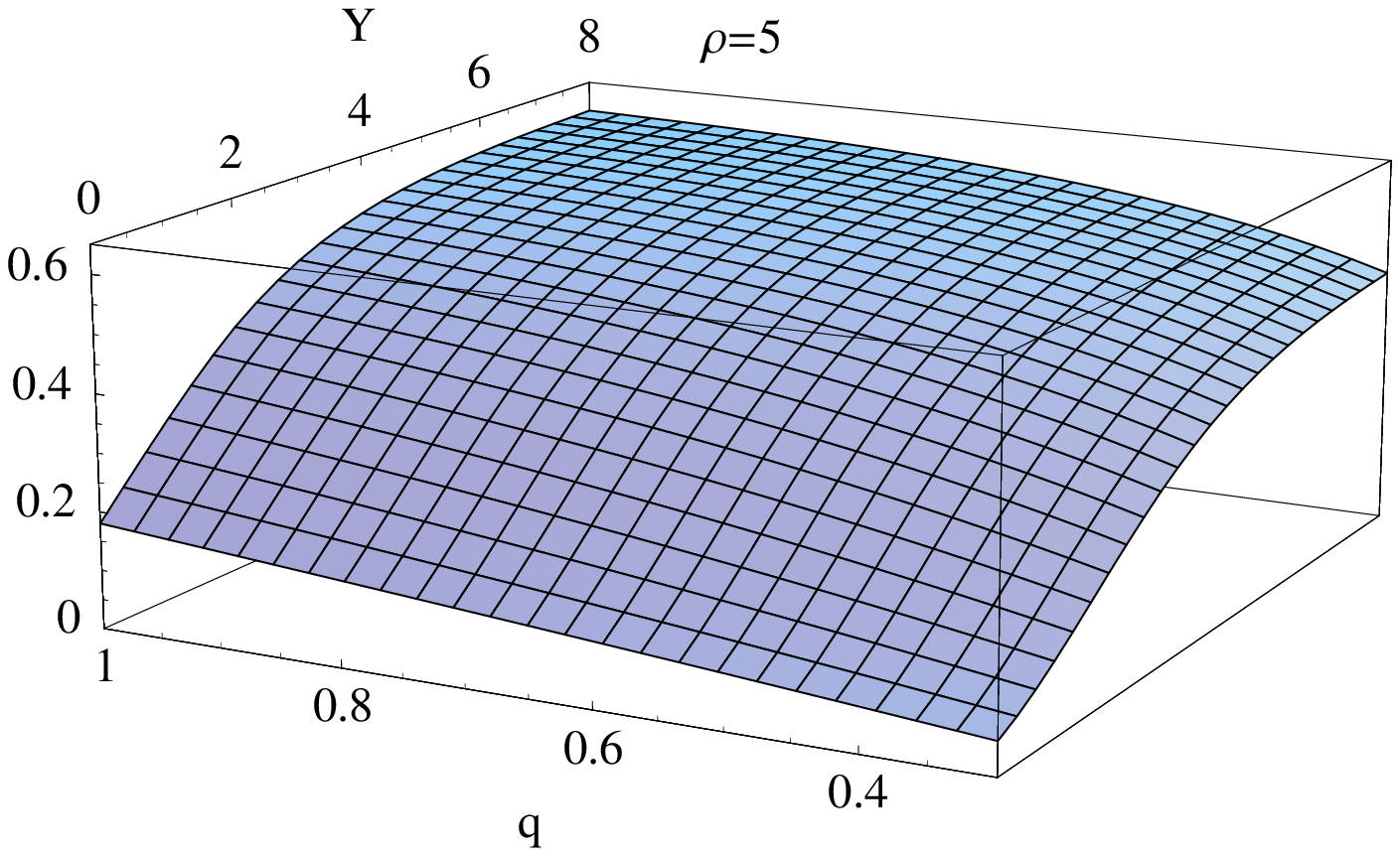,width=115mm} & 
\psfig{file=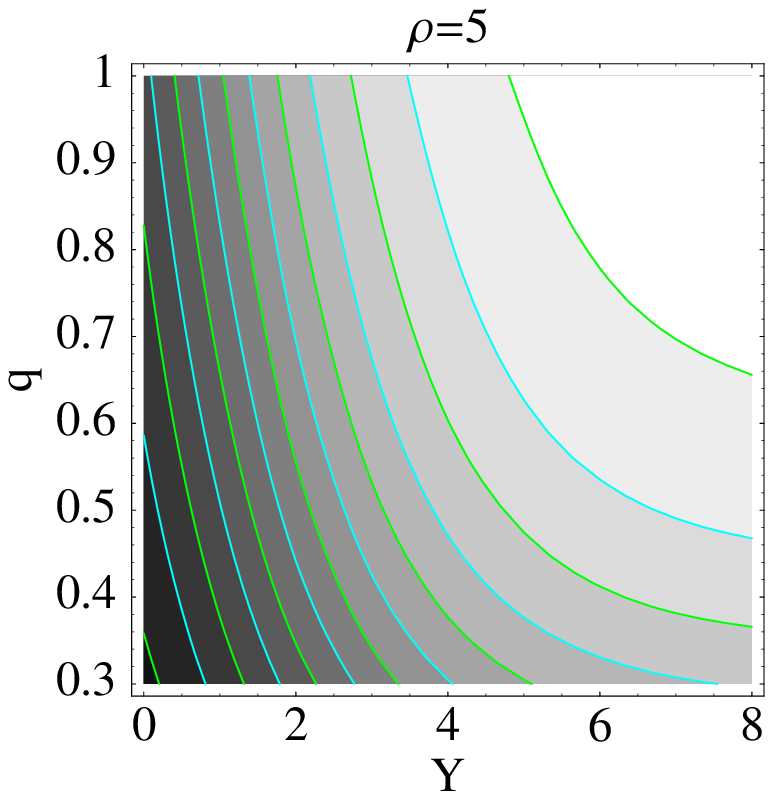,width=60mm}\\
 &  \\ 
\fig{EFUN66}-a & \fig{EFUN66}-b \\
 &  \\
\end{tabular}
\caption{\it
The quantum amplitude  $\Psi(Y,\,q)\,$ of the 3P vertex model
for the non symmetrical case
is represented by 3D plot and contour plot. The plots are presented as a functions
of scaled rapidity Y at $\ro\,=\,5\,$ and at fixed $q_1=0.2$ with  $q_2=0.3-1$.}
\label{EFUN66}
\end{figure}

\begin{figure}[hptb]
\begin{tabular}{ c c}
\psfig{file=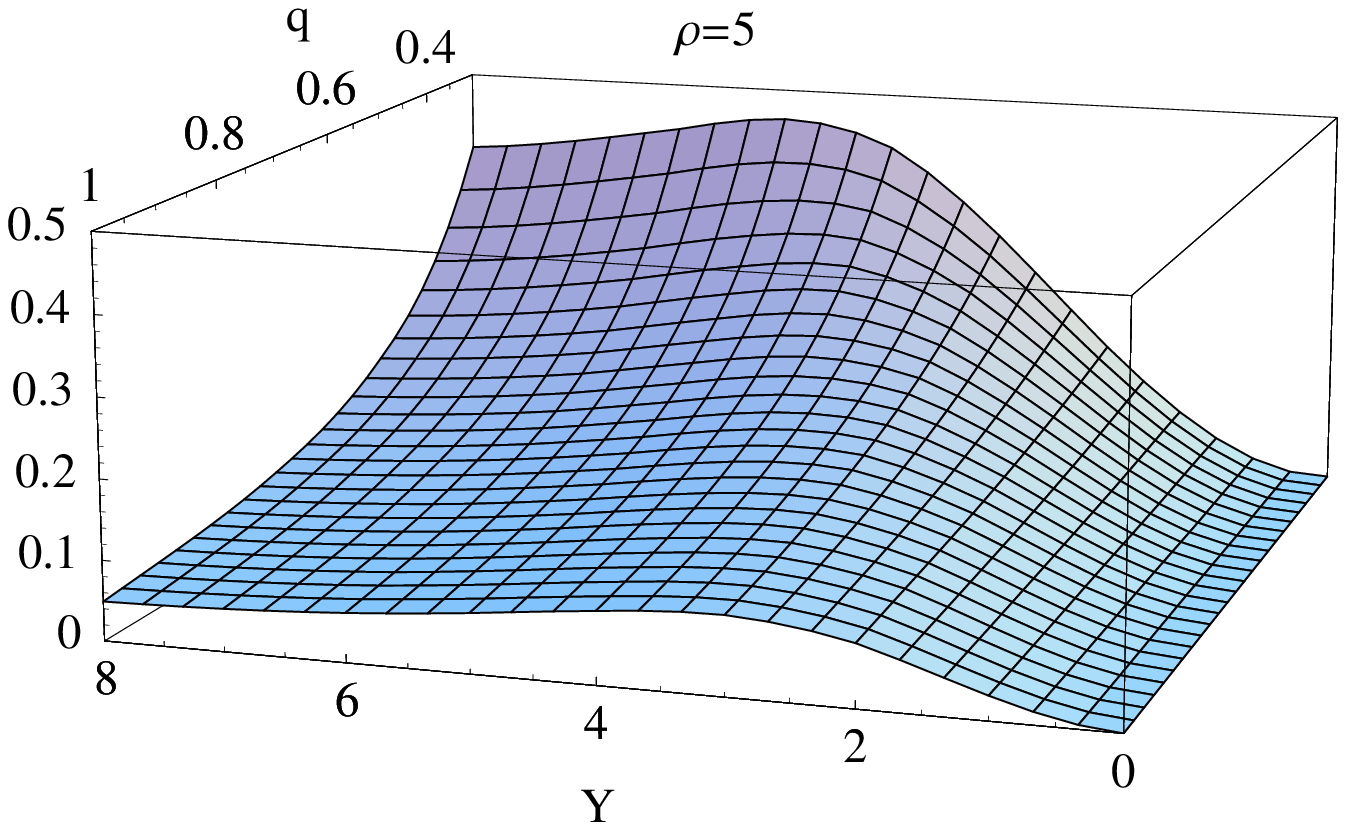,width=115mm} & 
\psfig{file=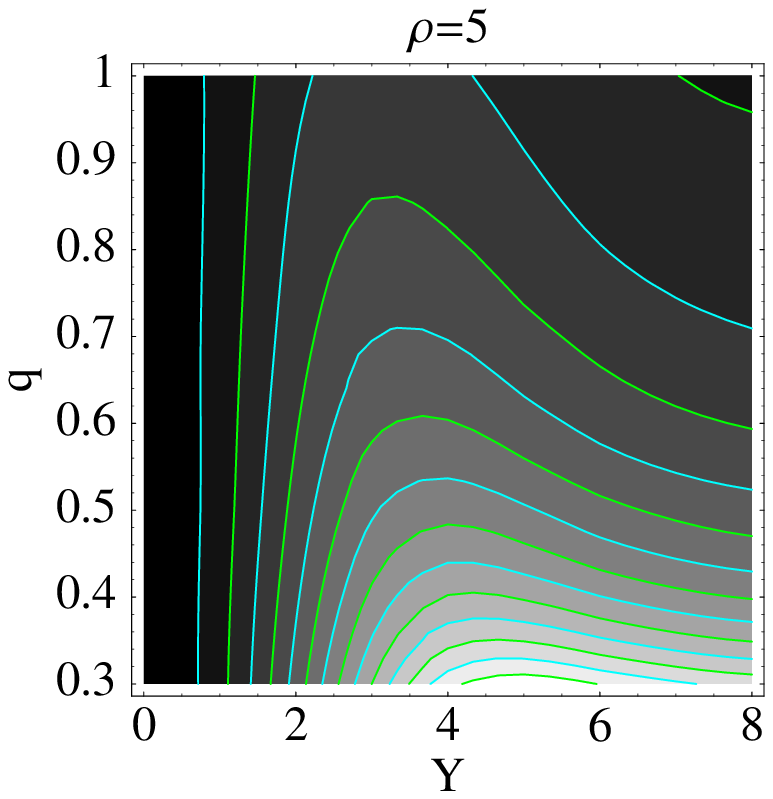,width=60mm} \\
 &  \\ 
\fig{EFUN666}-a & \fig{EFUN666}-b \\
 &  \\
\end{tabular}
\caption{\it 
The ratio $|\Psi(Y,\,q)\,-\,\Psi_{c}(Y,\,q_{c},\,p_{c})|\,/\,\Psi(Y,\,q)$
in the 3P vertex model
is represented by 3D plot and contour plot
at  the non symmetrical values of the 
sources:  $q_1=0.2\,$ is fixed and  $q_2=0.3-1$ is a variable.
The plots are presented as a functions
of scaled rapidity Y at $\ro\,=\,5\,$ . }
\label{EFUN666}
\end{figure}

  The results of the calculations 
of the scattering amplitude $iT$ for the quantum and classical cases
are presented in the Fig.\ref{EFUN3}-Fig.\ref{EFUN6}
for the symmetrical values of the external sources.
In the Fig.\ref{EFUN66}-Fig.\ref{EFUN666} the results for 
the non symmetrical case are represented. 
These results are presented in the form of 3D plots as well, as in the form
of contour plots. The white color of the contour plot corresponds
to the maximum value of the z axis function of the 3D plot, and the  black 
color , correspondingly, to the minimum value of the same function. 
The amplitude is given as a function 
of two variables, rapidity and value of sources.
In a symmetrical case the sources are the same, in non symmetrical case
we fix one sources, $q_1\,=\,0.2\,$, and the second variable
is the value of the second, external source $q_2\,=\,0.3\,-\,1\,$.
We also present the absolute ratio of the difference
between quantum amplitude and classical amplitude to the quantum amplitude,
$|\Psi(Y,\,q)\,-\,\Psi_{c}(Y,\,q_{c},\,p_{c})|\,/\,\Psi(Y,\,q)$, 
in order to illustrate the relative contribution of the loops in the full solution 
comparatively to the classical solution.
The rapidity variable Y of the plots is scaled, i.e. 
in the plots the variable  $\,\mu\,y\,$ was used
instead usual rapidity $y$.
Therefore, in order
to obtain the value of the amplitude at some physical rapidity
it needs to divide the Y variable of the plot on
intercept $\mu$ : $Y_{phys}\,=\,Y_{plot}\,/\,\mu\,$. For any
value of intercept $\mu$ we have the plot of the amplitude
for rapidities $Y_{phys}\,=\,Y_{plot}\,/\,\mu\,$ 
at triple Pomeron vertex $\lambda$ equal to 
$\lambda\,=\,\mu\,/\,\ro\,\,$.
We see, that for the maximum value
of the $Y_{plot}\,=\,8\,$ at intercept, for example, 
$\,\mu\,=\,0.3\,$ the normal rapidity is large: $\,Y_{phys}\,\sim\,27\,$
and covers the reasonable rapidity range which may be interesting in the high energy physics.

\begin{figure}[hptb]
\begin{tabular}{ c c}
\psfig{file=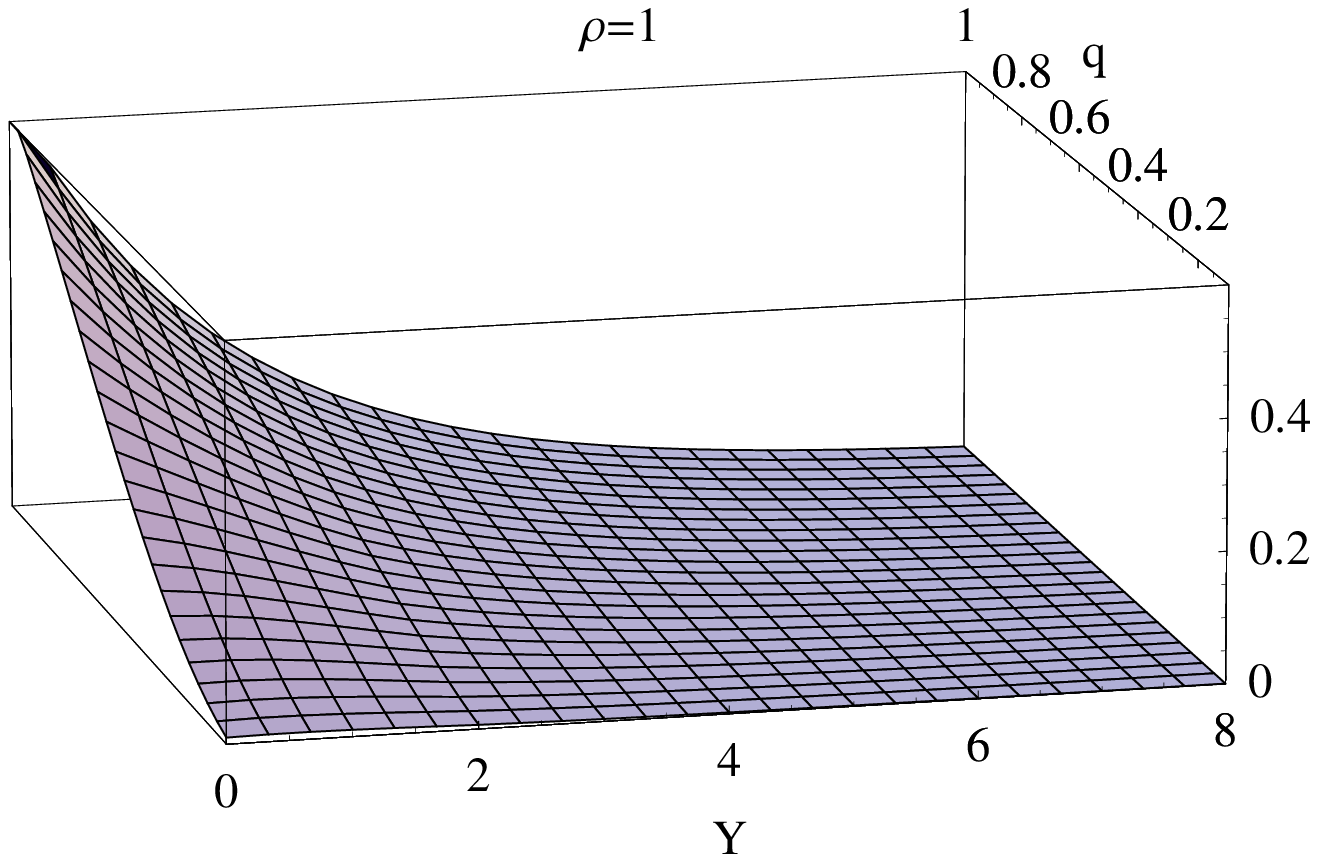,width=120mm} & 
\psfig{file=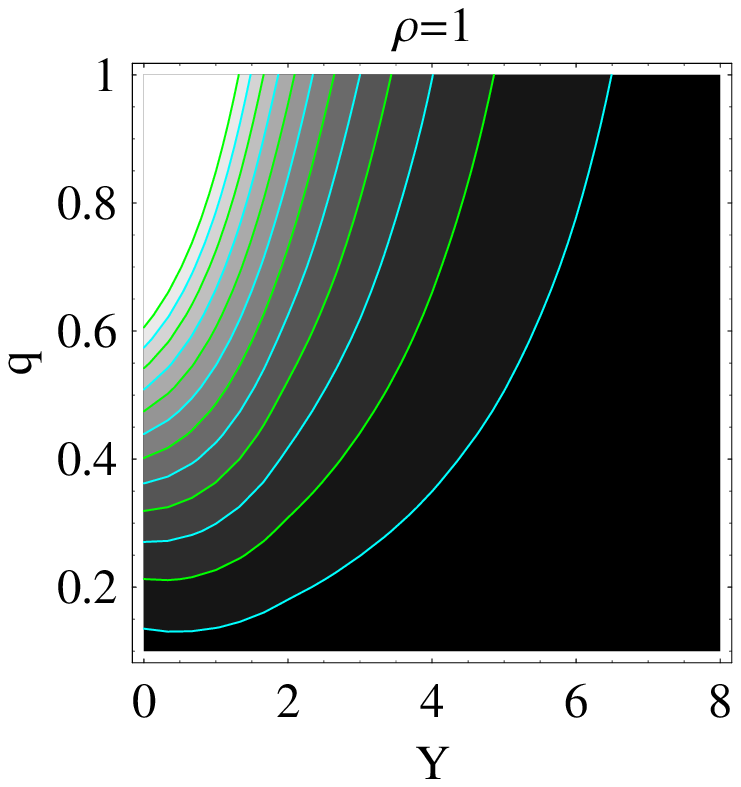 ,width=55mm}\\
 &  \\ 
\fig{EFUN7}-a & \fig{EFUN7}-b \\
 &  \\
\end{tabular}
\caption{\it The 
quantum amplitude  $\Psi(Y,\,q)\,$ of the 3P vertex model
in the form of 3D and contour plots
at $\ro\,=\,1\,$  as a functions of scaled
rapidity Y and symmetrical values of external sources $q_1=q_2$.}
\label{EFUN7}
\end{figure}

  We also present a quantum amplitude  calculated
for the small  value of $\ro\,=\,1$.
This value of $\ro\,$ means large value of triple Pomeron vertex
and relatively large value of the ground state energy, see table
\ref{EFUN}.
From the plot Fig\ref{EFUN7} is clearly seen, that at large enough 
rapidities the amplitude approaches
zero as it must be for such value of $\ro\,$.

%%%%%%%%%%%%%%%%%%%%%%%%%%%%%%%%%%%%%%%%%%

\subsection{''Effective'' Pomeron propagator}

 With the knowledge of the spectrum of RFT, the calculation
of the  ''effective'' Pomeron propagator $P_{eff}(y,q)$ is easy task.
First of all, we change the initial condition for our equations,
instead Eq.\ref{Quant6} now we have

\beq\label{Prop1}
\Psi(y=0,q)\,=\,I(q)\,=\,I(q\,,q_{ext})\,=\,q\,q_{ext}\,.
\eeq
The scattering amplitude $\Psi(y,q)$ now describes the Green's function
of transition from one Pomeron created from
source $q$ at rapidity zero 
to any number N  Pomerons interacting with 
sources $q_{ext}$ at final rapidity.

\begin{figure}[hptb]
\begin{center}
\begin{tabular}{ c c}
\psfig{file=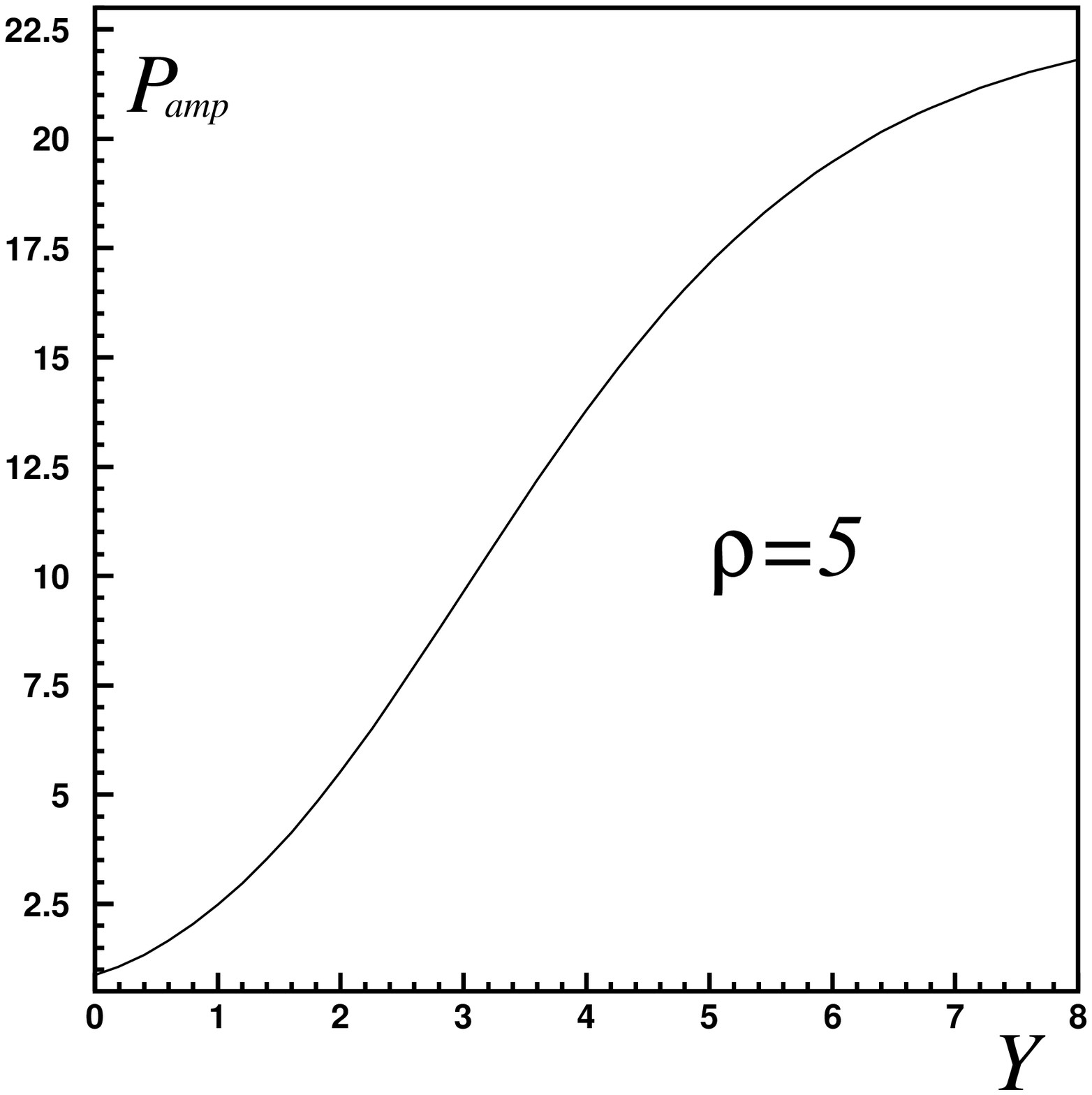,width=80mm} & 
\psfig{file=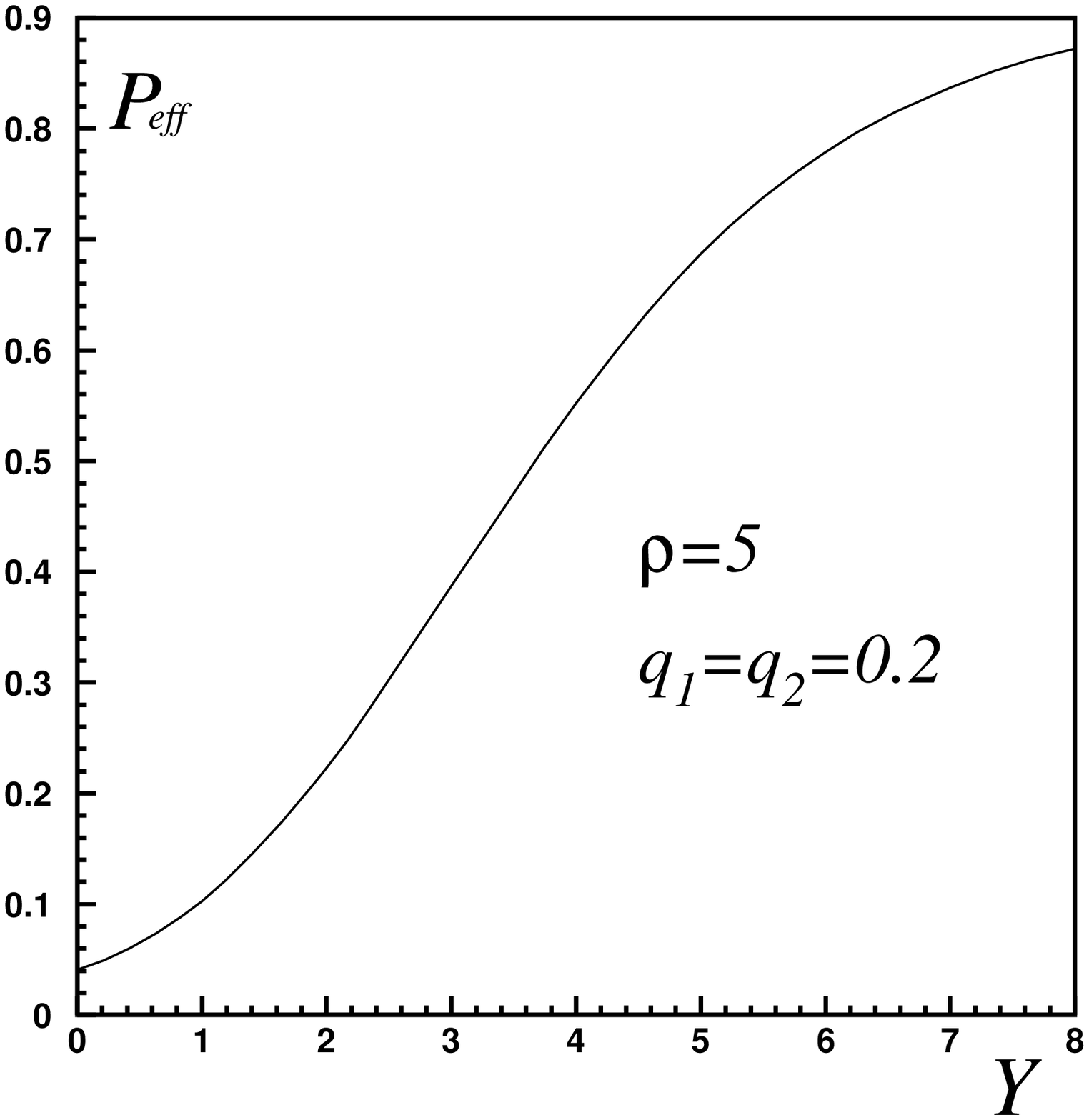,width=80mm}\\
 &  \\ 
\fig{EFUN8}-a & \fig{EFUN8}-b \\
 &  \\
\end{tabular}
\end{center}
\caption{\it The  $P_{amp}(y,q)$ and $P_{eff}(y,q)$ 
functions of the 3P model
at $\ro\,=\,5\,$  and 
$q_1\,=\,q_{2}\,=1/ \ro\,=0.2\,$ as a functions of scaled
rapidity Y .}
\label{EFUN8}
\end{figure}
\begin{figure}[hptb]
\begin{tabular}{ c c}
\psfig{file=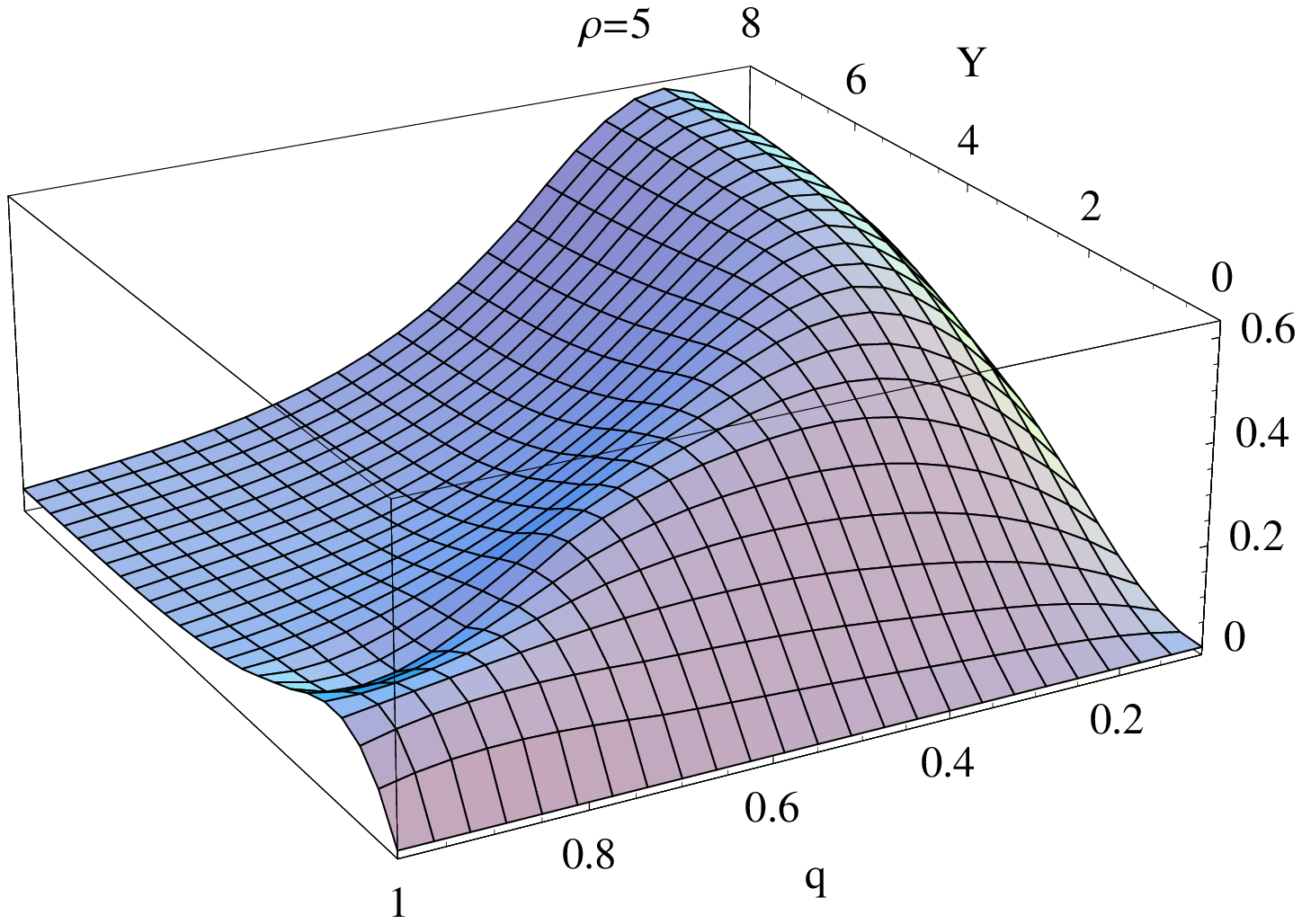,width=115mm} & 
\psfig{file=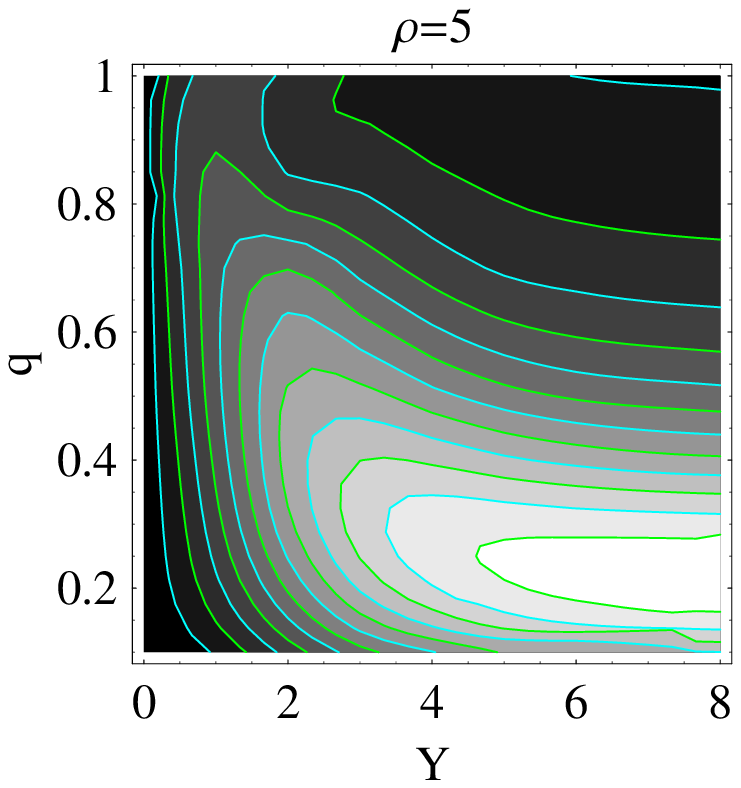,width=60mm}\\
 &  \\ 
\fig{EFUN10}-a & \fig{EFUN10}-b \\
 &  \\
\end{tabular}
\caption{\it The ratio 
$|\Psi(Y,\,q)\,-\,\Psi_{eff}^{eik}(y,q)\,|\,/\,\Psi(Y,\,q)$
in the 3P model
in the form of 3D and contour plotshe 3P model
at $\ro\,=\,5\,$  as a functions of scaled
rapidity Y and symmetrical values of external sources $q_1=q_2=0.1-0.2$.}
\label{EFUN10}
\end{figure}

Therefore, in order to obtain requested "effective" Pomeron propagator
with only one Pomeron interacting with one source at final rapidity,
we also need to take the derivative of  $\Psi(y,q)$
over Pomeron field $q$ at $\,q\,=\,0\,$
and multiply obtained function on Pomeron field $q$ :
\beq\label{Prop2} 
P_{eff}(y,q)\,=\,\Le\frac{d\,\Psi(y,q)}{d\,q}\Ra_{q=0}\,\cdot\,q\,\,.
\eeq
The $P_{eff}(y,q)$ function is simply the second term of the Tailor expansion  
of the $\Psi(y,q)$ around $q=0$. 
Dividing obtained propagator $P_{eff}(y,q)$ on the sources $q\,q_{ext}\,$,
we obtain the Green's function of the theory, which does not depend on the values of the sources:
\beq\label{Prop22} 
P_{amp}(y,q)\,=\,\frac{P_{eff}(y,q)}{q\,q_{ext}\,}.
\eeq
$P_{amp}(y,q)$ does not depend on the sources of the Pomerons fields 
and in order to obtain requested propagator with given sources
we need simply to multiply $P_{amp}(y,q)$ on the values of these sources
$q\,q_{ext}\,$. Fig\ref{EFUN8} presents a plot of the  $P_{amp}(y,q)$ function
and plot for the $P_{eff}(y,q)$ function at $q\,=\,q_{ext}\,=1/ \ro\,=0.2\,$.
The interesting question, 
which we can address in this calculations, it is the question about the importance of $P_{eff}(y,q)$ in  RFT-0. 
Namely, let's define with the help of $P_{eff}(y,q)$  the eikonalized ''effective''
Pomeron amplitude
\beq
\Psi_{eff}^{eik}(y,q)\,=\,1-e^{-P_{eff}(y,q)}\,
\label{EikProp}
\eeq
which account loops in the "effective" propagator
and does not account interactions between different propagators
and between propagators and external sources.
This amplitude could clarify the precision of this approximation in comparison with the full solution.
So, the required ratio is presented in the Fig\ref{EFUN10} .

%%%%%%%%%%%%%%%%%%%%%%%%%%%%%%%%%%%%%%%%%%

\section{Solution of RFT in zero transverse dimension with triple and 
quaternary Pomeron vertices}

 In this section we discuss a
RFT-0 model given by the following Lagrangian:
\beq\label{FV1}
L\,=\,q\,\dot{p}\,+\,\mu\,q\,p\,-\,\lambda\,q\,(q\,+\,p)\,p\,+
\,\lambda^{'}\,q^2\,p^2\,,
\eeq 
where new quaternary Pomeron vertex $\,\lambda^{'}\,$ is introduced in 
comparison to the Eq.\ref{Lag2}.
In the terms of the $q,\,p\,$ operators, see  Eq.\ref{Oper},
the Hamiltonian of the problem has the form
\beq\label{FV2}
H=\,(\mu\,q\,-\,\lambda\,q^{2})\,\frac{d}{dq}\,+
\,(\lambda\,q\,-\,\lambda^{'}\,q^{2})\,\frac{d^2}{dq^2}\,\,,
\eeq
and the full, quantum solution of the theory
will coincide with the solution of the quantum mechanical problem
determined by the new Hamiltonian  Eq.\ref{FV2}:
\beq\label{FV3}
H\,\Psi\,=\,\frac{d\,\Psi}{d\,y}\,\,,
\eeq
where the function
\beq
\Psi(y,q)\,=\,\sum_{n=0}^{n=\infty}\,\lambda_{n}\,
e^{\,-\,E_{n}\,y}\,\phi_{n}(q)\,\,
\eeq
is the full quantum amplitude
of two scattering particles. As before, $E_{n}$ and $\phi_{n}(q)$
denote eigenvalues and eigenfunctions of the Hamiltonian given 
by the
Eq.\ref{FV2}, coefficients $\lambda_{n}$ are the normalized projections 
coefficients of the eigenfunctions on the value of $\,\Psi\,$ at $\,y=0\,$. 

  The 
value of the quaternary vertex $\,\lambda^{'}\,$ is important for our following calculations. 
If the vertex is very small,
$\,\lambda^{'}\,<\,\lambda\,/\,\ro\,$ then all results of the Section 1
will stay the same with only very small corrections to the amplitude, it is was known many years
ago, see  \cite{0dimq}. Our calculations support this conclusion.
Therefore, for our calculations more interesting to take another
value of the vertex, so called ''magic'' value of $\,\lambda^{'}\,$, which is given by the following expression
\beq\label{FV33}
\lambda^{'}\,=\,\frac{\lambda}{\ro}\,.
\eeq
The important property of the theory with this ''magic'' value of 
$\,\lambda^{'}\,$ is that in this theory , unlike previous cases, 
the ground state is precisely zero, as we will see later.
Another important feature of the Hamiltonian Eq.\ref{FV2}
with the ''magic'' value of the vertex 
is that this model has precise correspondence
with the s-channel reaction-diffusion models, see \cite{BMMSX}, and with conformal
approach for the QCD Pomeron, see \cite{Korch,BondLang}.

   As in the previous model,  in this section we will also consider only the
calculations with the 
value of the parameter  $\ro\,=\,5$.
Nevertheless, the qualitative understanding of the behavior of the amplitude
at smaller values of $\ro$ also will be clear.

%%%%%%%%%%%%%%%%%%%%%%%%%%%%%%%%%%%%%%%%%%

\subsection{The quantum solution of the second model}

 The quantum solution
of the Hamiltonian Eq.\ref{FV2} is a solution of 
the following second order differential equation 
\beq\label{QFV1}
\,\lambda^{'}\,q\,(\,\lambda\,/\,\lambda^{'}\,-\,q\,)\,
\,\frac{d^2\,\Psi(y,q)}{dq^2}\,+\,\,\lambda\,q\,(
\ro\,-\,q)\,\frac{d\,\Psi(y,q)}{dq}\,=\,\frac{d\,\Psi(y,q)}{d\,y}\,\,
\eeq
with the initial and boundary conditions on the function $\Psi(y,q)$:
\begin{eqnarray}\label{QFV2}
\,&\,&\,\Psi(y=0,q)\,=\,\sum_{n=0}^{n=\infty}\,\lambda_{n}\,
\,\phi_{n}(q)\,=\,I(q)\,\\
\,&\,&\,\Psi(y,q\,\rightarrow\,0)\,\propto\,q\,\\
\,&\,&\,\Psi(y,q\,\rightarrow\,\ro)\,\propto \,const.\,\,,
\end{eqnarray}  
where as before, the form of the  function $I(q)$ depends on the particular physical problem.
Very important property of Eq.\ref{QFV1} is the existing of the ground state
$\phi_{0}(q)$ with the zero energy $E_0\,=\,0\,$.
In this case the solution for Eq.\ref{QFV1} is trivial:
\beq\label{QFV3}
\phi_{0}(q)\,=\,1\,-\,e^{-\ro\,q}\,.
\eeq
This ground state is not orthogonal to the other
eigenfunctions of the Hamiltonian, $\phi_{i},\,\,i=\,1\,..\,\infty\,$,
which are orthogonal to each other:
\beq\label{QFV4}
\int_{0}^{\ro}\,\phi_{i}(q)\,\phi_{j}(q)\,F_{W}(q,\ro)\,dq\,=
\,Const\,\delta_{i\,j}\,
\eeq
with the weight function 
\beq
F_{W}(q,\ro)\,=\,\frac{e^{\ro\,q}}{q\,(\ro\,-\,q)}\,.
\eeq
where
\begin{eqnarray}\label{QFV55}
\,&\,&\,\phi_{i}(q=0)\,=\,0,\,\,\,\,i=\,1\,..\,\infty\,;\\
\,&\,&\,\phi_{i}(q=\ro)\,=\,0,\,\,\,\,i=\,1\,..\,\infty\,.
\end{eqnarray} 
Using these properties of th eigenfunctions we obtain for the
$\,\lambda_0$ projection coefficient:
\beq\label{QFV5}   
\lambda_0\,=\,\frac{I(\ro)}{\phi_{0}(\ro)}\,=\,\frac{I(\ro\,,q_{ext})}
{\,1\,-\,e^{-\ro^2}\,}
\eeq
All other projection coefficients, $\lambda_{i},\,\,i=\,1\,..\,\infty\,$,
may be calculated using the following expression:
\beq\label{QFV6}   
\lambda_{i}(q_{ext})\,=\,\frac{\int_{0}^{\ro}\,\phi_{i}(q)\,I(q\,,q_{ext})\,
F_{W}(q,\ro)\,dq\,
}{\int_{0}^{\ro}\,\phi_{i}^{2}(q)\,F_{W}(q,\ro)\,dq\,}\,-\,
\lambda_0\,\frac{\int_{0}^{\ro}\,\phi_{i}(q)\,\phi_{0}(q)\,
F_{W}(q,\ro)\,dq\,
}{\int_{0}^{\ro}\,\phi_{i}^{2}(q)\,F_{W}(q,\ro)\,dq\,}\,.
\eeq
For the functions $\phi_{i},\,\,i=\,1\,..\,\infty\,$ the equation Eq.\ref{QFV1}
has the form
\beq\label{QFV7} 
\,\frac{d^2\,\phi_i(q)}{dq^2}\,+\,\ro\,
\frac{d\,\phi_i(q)}{dq}\,=\,-\,\frac{E_i}{\lambda^{'}\,q\,(\ro-q)}\,\phi_i(q)\,,
\eeq
and with the help of the transformation
\beq\label{QFV8} 
\phi_i(q)\,=\,f_i(q)\,e^{-\ro\,q\,/\,2}
\eeq
this equation gets the following standard Shredinger form:
\beq\label{QFV9}  
\,\frac{d^2\,f_i(q)}{dq^2}\,-\,\frac{\ro^2}{4}\,f_i(q)\,=\,-
\,\frac{E_i}{\lambda^{'}\,q\,(\ro-q)}\,f_i(q)\,
\eeq
for the $\,f_i(q)\,$ functions.
The function $f_i(q)$ are defined on the edges of interval as 
\begin{eqnarray}\label{QFV10}
\,&\,&\,f_{i}(q=0)\,=\,0,\,\,\,\,i=\,1\,..\,\infty\,;\\
\,&\,&\,f_{i}(q=\ro)\,=\,0,\,\,\,\,i=\,1\,..\,\infty\,,
\end{eqnarray} 
and they are orthogonal each to other on the interval from $\,0\,$
to $\,\ro\,$ with the new weight function $F_{W}^{f}(q,\ro)$
\beq
F_{W}^{f}(q,\ro)\,=\,\frac{1}{q\,(\ro\,-\,q)}\,.
\eeq
Our problem we also can solve using hermitian 
Hamiltonian defined by Eq.\ref{QFV9} with the projection coefficients
\beq\label{QFV66}   
\lambda_{i}(q_{ext})\,=\,\frac{\int_{0}^{\ro}\,f_{i}(q)\,I(q\,,q_{ext})\,
\,e^{\ro\,q\,/\,2}\,F_{W}^{f}(q,\ro)\,dq\,
}{\int_{0}^{\ro}\,f_{i}^{2}(q)\,F_{W}^{f}(q,\ro)\,dq\,}\,-\,
\lambda_0\,\frac{\int_{0}^{\ro}\,f_{i}(q)\,\phi_{0}(q)\,
\,e^{\ro\,q\,/\,2}\,F_{W}^{f}(q,\ro)\,dq\,
}{\int_{0}^{\ro}\,f_{i}^{2}(q)\,F_{W}^{f}(q,\ro)\,dq\,}\,.
\eeq

 As was done in the previous section, solving the two value boundary problem
for differential equation Eq.\ref{QFV7} or Eq.\ref{QFV9},
we obtain the spectrum and eigenfunctions of the model and find
an amplitude for the different values of the sources $q_1,\,q_2\,$
and parameter $\ro$:
\beq\label{QFV11}
\Psi(y,q=q_2)\,=\,\frac{I(\ro\,,q_{ext})\,(1\,-\,e^{-\ro\,q}\,)\,}
{\,1\,-\,e^{-\ro^2}\,}\,+\,
\,\sum_{n=1}^{n=\infty}\,\lambda_{n}(q_1)\,
e^{-E_{n}\,y}\,\phi_{n}(q_2)\,\,.
\eeq  

%%%%%%%%%%%%%%%%%%%%%%%%%%%%%%%%%%%%%%%%%%

\subsection{The classical solution for the second model}

\begin{figure}[t]
\begin{center}
\psfig{file=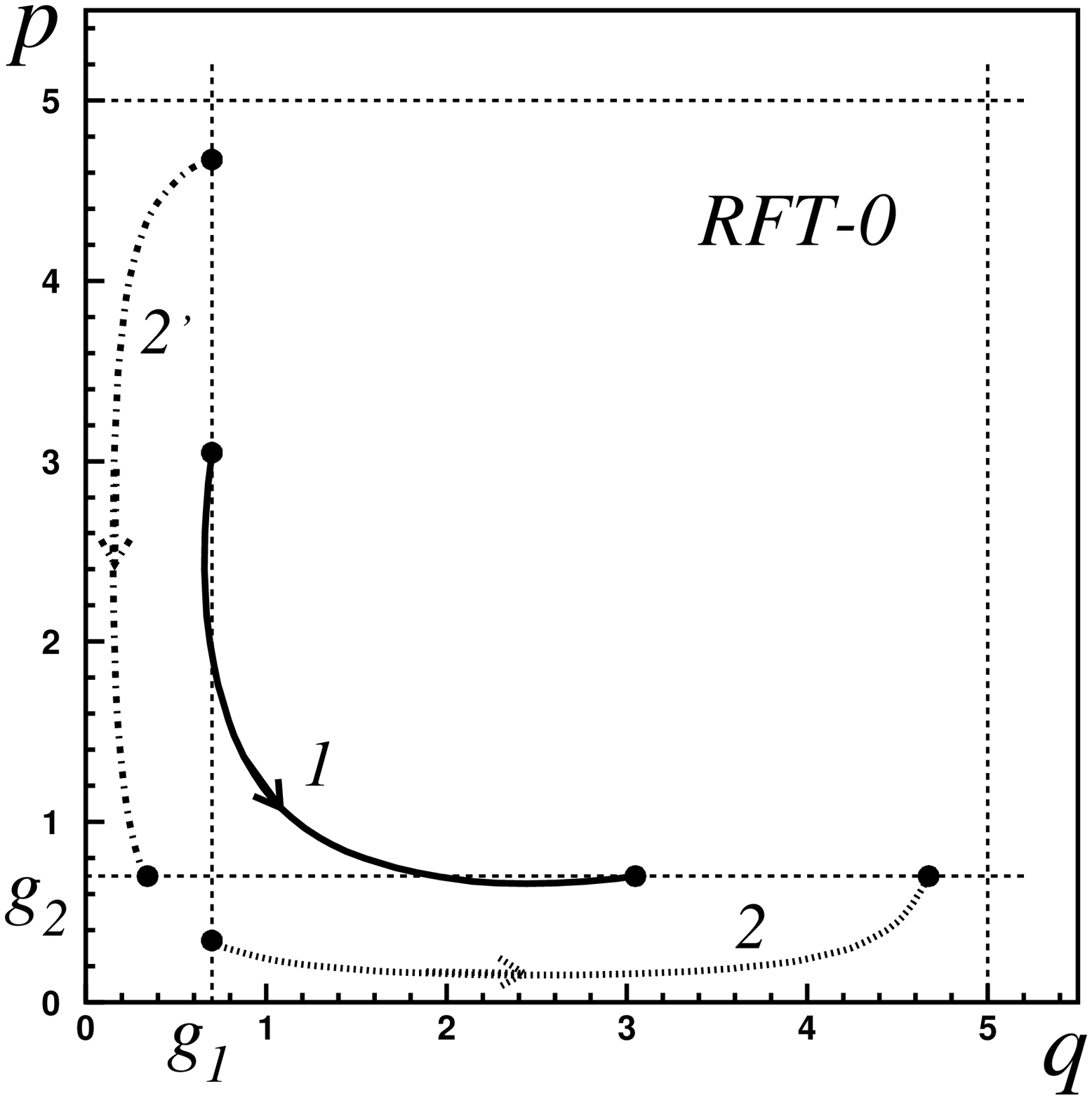,width=100mm} 
\end{center} 
\caption{\it Classical solutions of the RFT-0 with
triple and quaternary Pomeron vertexes: 
 the $\{ q,p\}$ trajectories obtained for $Y=5\,>\,Y_c\,$, $q=p=0.7$, 
$\rho\,=\,5$.}
\label{Traj4}
\end{figure}

 Comparison between the contribution to the amplitude of diagrams with loops and without loops
may be performed again if we additionally will calculate classical solutions 
for the Lagrangian Eq.\ref{FV1}: 
\beq\label{CFV1}
L\,=\,\frac{1}{2}\,q\,\dot{p}\,-\frac{1}{2}\,\dot{q}\,p\,+
\,\mu\,q\,p\,-\,\lambda\,q\,(q\,+\,p)\,p\,+\,
\,\lambda^{'}\,q^2\,p^2\,+\,
\,q(y)\,p_0(y)\,+\,q_0(y)\,p(y)\,\,.
\eeq
The equation of motion for the fields p and q 
in this case are the following:
\begin{eqnarray}\label{CVF2}
\,&\,&\,\dot{q}\,=\,\mu\,q\,-\,\lambda\,q^2\,-\,2\,\lambda\,q\,p\,+\,2\,
\lambda^{'}\,q\,p^2\,\\
\,&\,&\,\dot{p}\,=\,-\,\mu\,p\,+\,\lambda\,p^2\,+\,2\,\lambda\,q\,p\,-\,2\,
\lambda^{'}\,q^2\,p\,\\
\,&\,&\,q_0(y)\,=\,q_1\,\delta(y)\,\\
\,&\,&\,p_0(y)\,=\,q_2\,\delta(y-Y)\,\,.
\end{eqnarray}
The solutions of these equations, of course, are different
from the solutions of the equations of motion for the Lagrangian 
Eq.\ref{Class1}, but the representation of the amplitude $\Psi_{c}(Y)$
in the terms of three classical solutions $\{q^{i}_{c},p^{i}_{c}\}$ of the system
Eq.\ref{CVF2} is the same as in the Subsection 2.2 ( see also
Fig.\ref{Traj4}) :
\beq\label{CVF3}
\Psi_{c}(Y)=1+exp\{-S(Y,\,q^{1}_{c}(y)\,,p^{1}_{c}(y))\}-
exp\{-S(Y,\,q^{2}_{c}(y)\,,p^{2}_{c}(y))\}-
exp\{-S(Y,\,q^{3}_{c}(y)\,,p^{3}_{c}(y))\}\,.
\eeq
Therefore, here
we will not repeat all steps of the  Subsection 2.2.
The picture for the classical solution derived for the 
model with only triple Pomeron vertex is the same for the model
with both triple and quaternary vertices. The only difference between two models
is the value of critical rapidity $Y_c$ from which the additional
solutions arising, breaking the target-projectile symmetry of the initial 
symmetrical solution, but for our consideration it is not important.

%%%%%%%%%%%%%%%%%%%%%%%%%%%%%%%%%%%%%%%%%%

\subsection{Parameters and asymptotic behavior of the amplitude in the second model}

  As in  the previous case, for the second model we will consider
quantum amplitude
$\,\Psi(Y,\,q)\,$  for the following value of 
$\ro\,=\,\mu\,/\,\lambda\,$:
\begin{eqnarray}\label{CVF4}
\,&\,&\,\ro\,=\,5\,\,\,\,(\mu\,=\,0.2\,,\,\lambda\,=\,0.04\,);
\end{eqnarray}
and for the following values of the external sources (only symmetrical case ):
%\begin{itemize}\label{CVF5}
%\item 
\beq\label{CVF5}
\,q_1\,=\,q_2\,=\,0.1\,-\,1\,.
\eeq
%\end{itemize}
The ground state of this model has precisely zero eigenvalue, as we
showed before, 
see  Eq.\ref{QFV3} and for other eigenvalues of the model see Table.\ref{QFV6}.
\begin{table}[t]
\begin{center}
\begin{tabular}{|c|c|c|c|c|c|c|c|c|c|c|}
\hline
\, & \,& \,& \, &\, &\, &\, &\, &\, &\, &\,\\
$\ro\,$ &  $E_0$ & $E_1$ & $E_2$ & $E_3$ &
 $E_4$ & $E_5$ & $E_6$ & $E_7$ & $E_8$ &
 $E_9$ \\
\, & \,& \,& \, &\, &\, &\, &\, &\, &\, & \\
\hline
\, & \,& \,& \, &\, &\, &\, &\, &\, &\, &\, \\
$\ro\,=\,5\,$ & 0 & 0.182 & 0.189 & 0.307 &
 0.338 & 0.42 & 0.513 & 0.626 & 0.756 & 0.905 \\
\, & \,& \,& \, &\, &\, &\, &\, &\, &\, &\,\\
\hline
\end{tabular}
\caption{\it The eigenvalues of Eq.\ref{QFV7} for the value
of the parameter $\ro$ equal 5.}
\label{CVF6}
\end{center}
\end{table}
Indeed, 
now, at asymptotically
large rapidity, the amplitude is fully defined by the ground state:
\beq\label{CVF7}
\Psi_{asymp.}(q)\,=\,\lambda_0\,\phi_{0}(q)\,=
\,\frac{I(\ro\,,q_{ext})\,(\,1\,-\,e^{-\ro\,q}\,)}
{\,1\,-\,e^{-\ro^2}\,}\,.
\eeq
In the case of the eikonal type initial conditions $I(q\,,q_{ext})$
the quantum amplitude $\Psi(y,q)\,$ at $q\,=\,\ro$ 
has the following form
\beq\label{CVF8}
I(\ro\,,q_{ext})\,=\,1\,-\,e^{-\ro\,q_{ext}}\,,
\eeq
and, therefore, we obtain the following asymptotic behavior of our quantum amplitude :
\beq\label{CVF9}
\Psi_{asymp.}(q)\,=
\,\frac{(1\,-\,e^{-\ro\,q_{ext}})\,(1\,-\,e^{-\ro\,q})}
{\,1\,-\,e^{-\ro^2}\,}\,.
\eeq
For the ''effective'' Pomeron propagator, $P_{eff}(y,q)$
the initial condition is different:
\beq\label{CVF10}
I(\ro\,,q_{ext})\,=\,\ro\,q_{ext}\,.
\eeq
Using the definition of the propagator given by Eq.\ref{Prop2} at
asymptotically large rapidity we obtain:  
\beq\label{CVF11}
P_{eff}^{asymp.}(q)\,=\,
\,\frac{(\ro\,q_{ext})\,(\ro\,q)}
{\,1\,-\,e^{-\ro^2}\,}\,.
\eeq
As we see, for each particular values of sources $q_1,q_2$ and parameter $\ro$
neither quantum amplitude nor full Pomeron propagator are not decreasing
to zero at asymptotically large rapidity, approaching instead some constant
values defined by  Eq.\ref{CVF9}  and Eq.\ref{CVF11}  correspondingly. 

%%%%%%%%%%%%%%%%%%%%%%%%%%%%%%%%%%%%%%%%%%

\subsection{Numerical results for the  the second model}

\begin{figure}[hptb]
\begin{tabular}{ c c}
\psfig{file=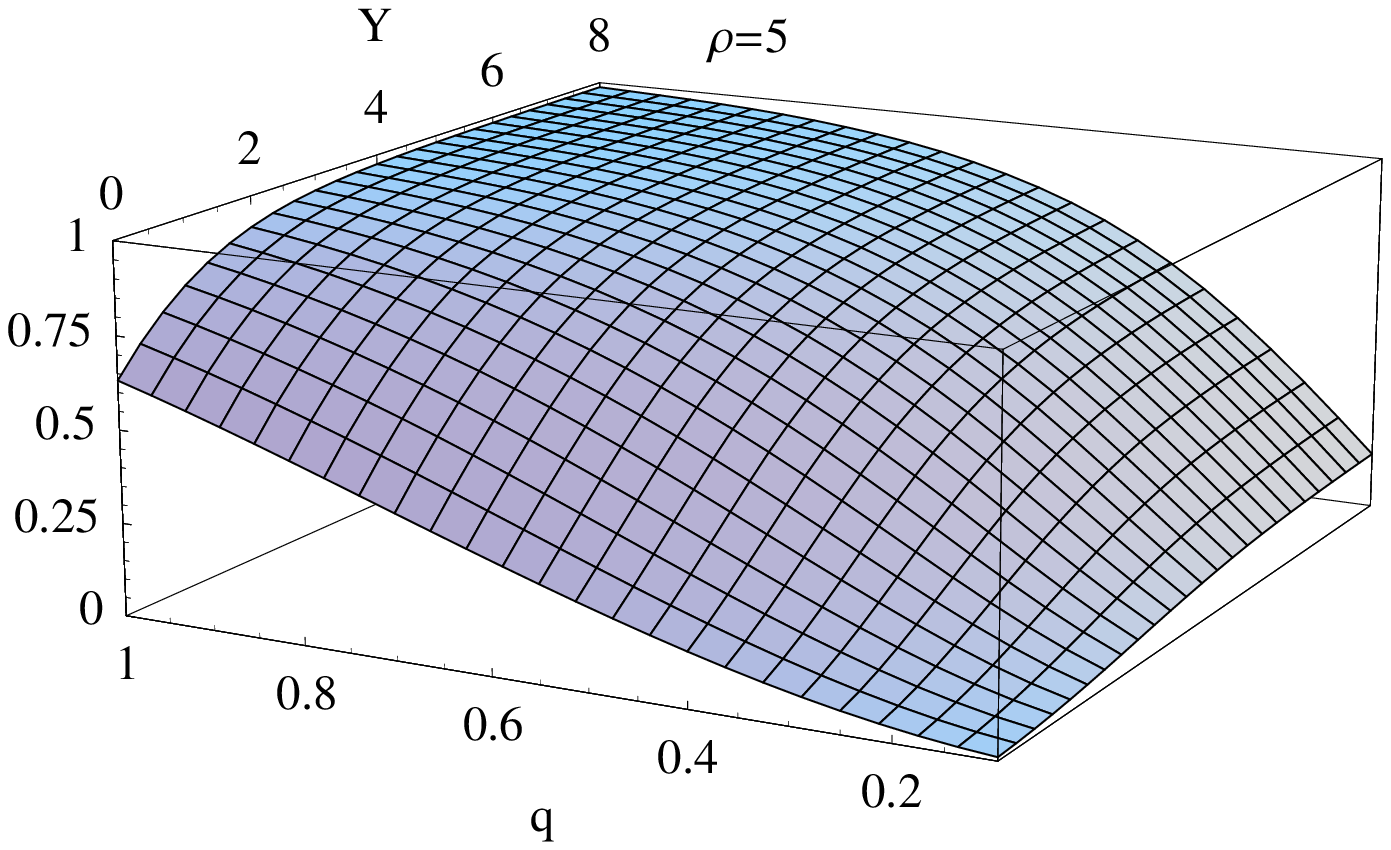 ,width=115mm} & 
\psfig{file=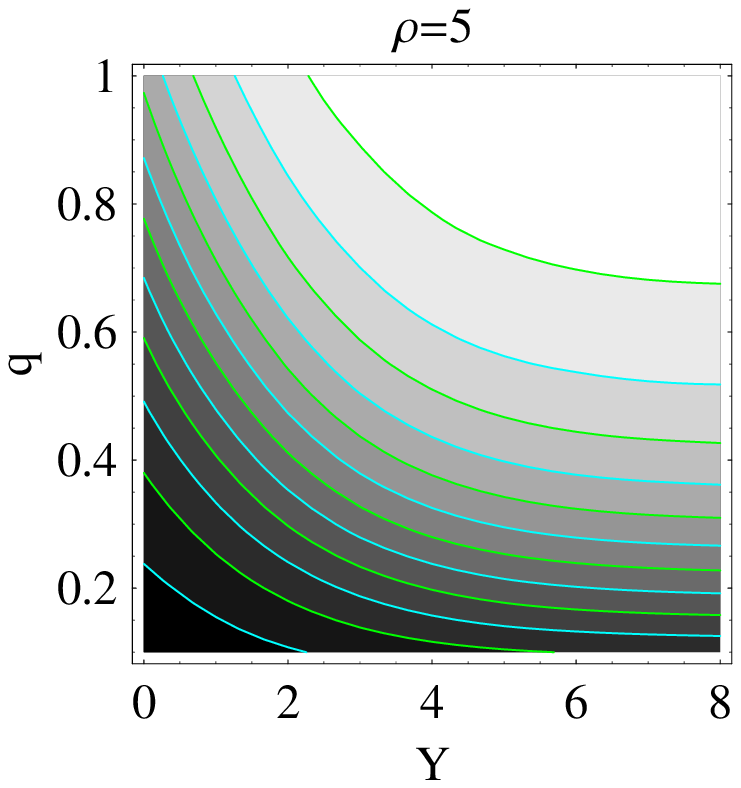 ,width=60mm}\\
 &  \\ 
\fig{NFV1}-a & \fig{NFV1}-b \\
 &  \\
\end{tabular}
\caption{\it 
The 
quantum amplitude  $\Psi(Y,\,q)\,$ of the second model
in the form of 3D and contour plots
at $\ro\,=\,5\,$  as a functions of scaled
rapidity Y and symmetrical values of external sources $q_1=q_2$.}
\label{NFV1}
\end{figure}

\begin{figure}[hptb]
\begin{center}
\psfig{file=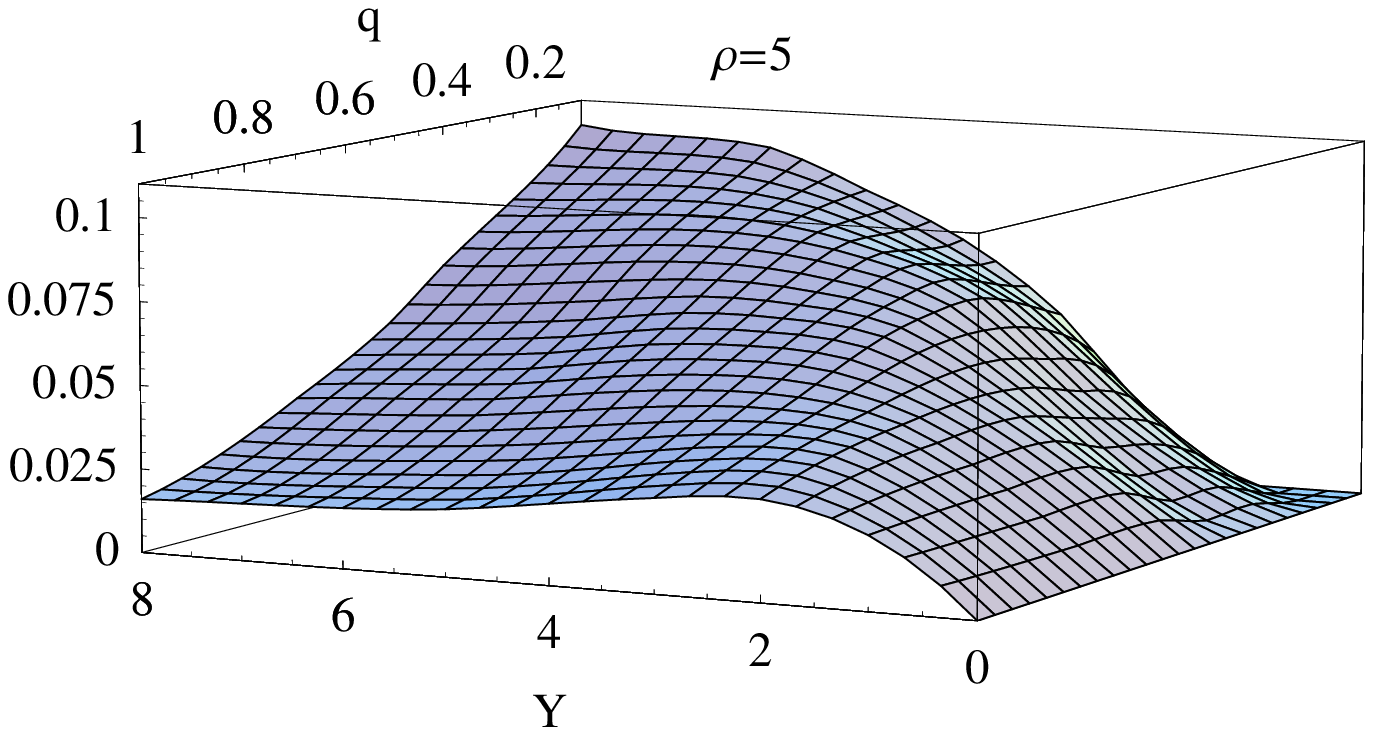,width=170mm} 
\end{center}
\caption{\it 
The ratio of the 
quantum amplitudes
$|\,\Psi^{3P}\,-\,\Psi^{4P}\,|\,/\,\Psi^{3P}$ 
in the second model
in the form of 3D  plot
at $\ro\,=\,5\,$  as a functions of scaled
rapidity Y and symmetrical values of external sources $q_1=q_2$.}
\label{NFV2}
\end{figure}

  The results of our calculations
of the scattering amplitude $iT$ for the quantum case
are presented in the same from as for the first model.
Due the similarities between these two models, in this section
we present only quantum solution for the amplitude
for the symmetrical case of interactions, see 
Fig.\ref{NFV1}. We also present
the ratio between the difference of the
triple Pomeron vertex model quantum amplitude and  
quantum amplitude of the second model to the the 
quantum amplitude of the first model :
$|\,\Psi^{3P}\,-\,\Psi^{4P}\,|\,/\,\Psi^{3P}$ ,
see Fig.\ref{NFV2}.

%%%%%%%%%%%%%%%%%%%%%%%%%%%%%%%%%%%%%%%%%%

\subsection{''Effective'' Pomeron propagator in the second model}

\begin{figure}[hptb]
\begin{tabular}{ c c}
\psfig{file=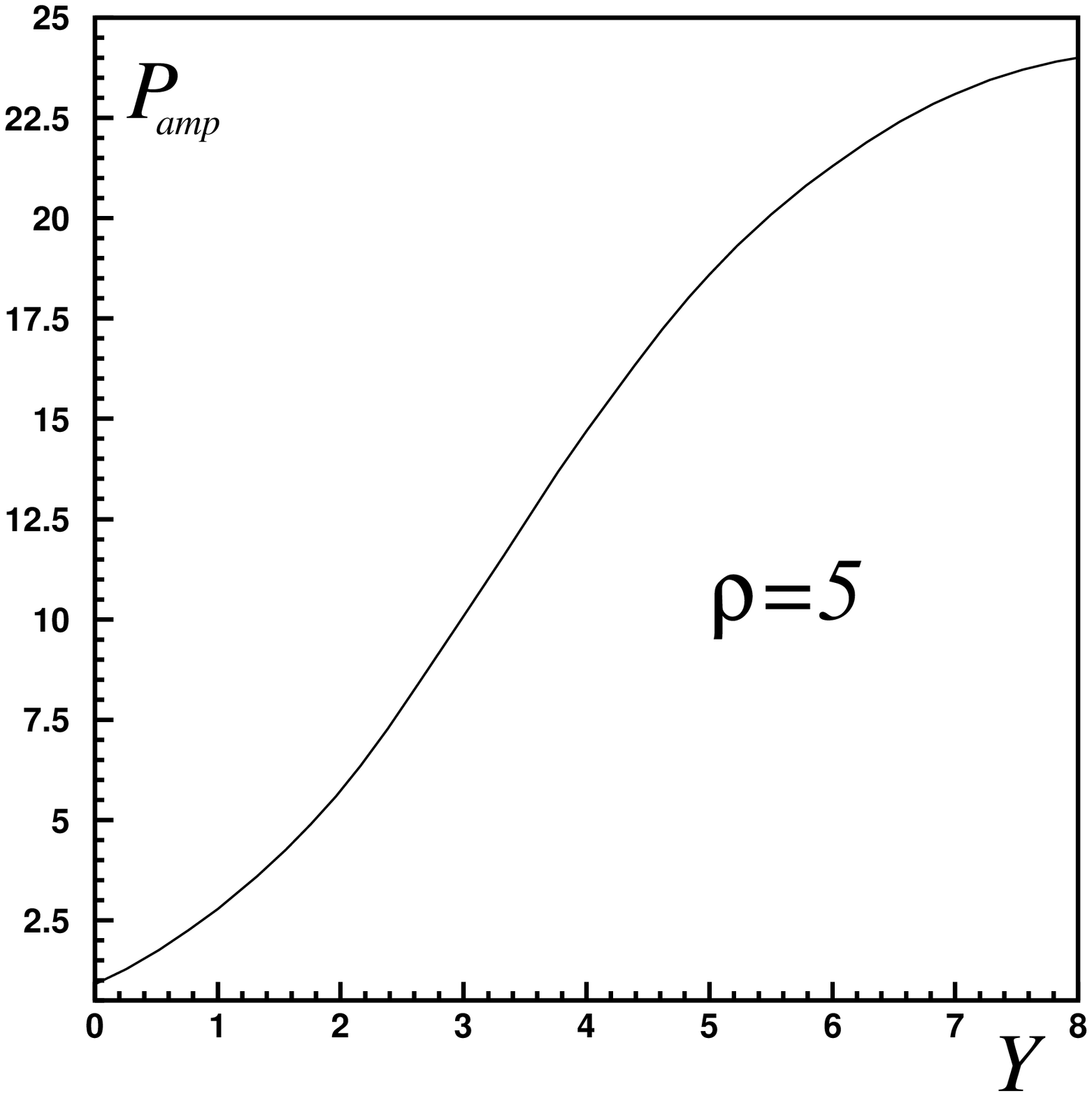,width=85mm} & 
\psfig{file=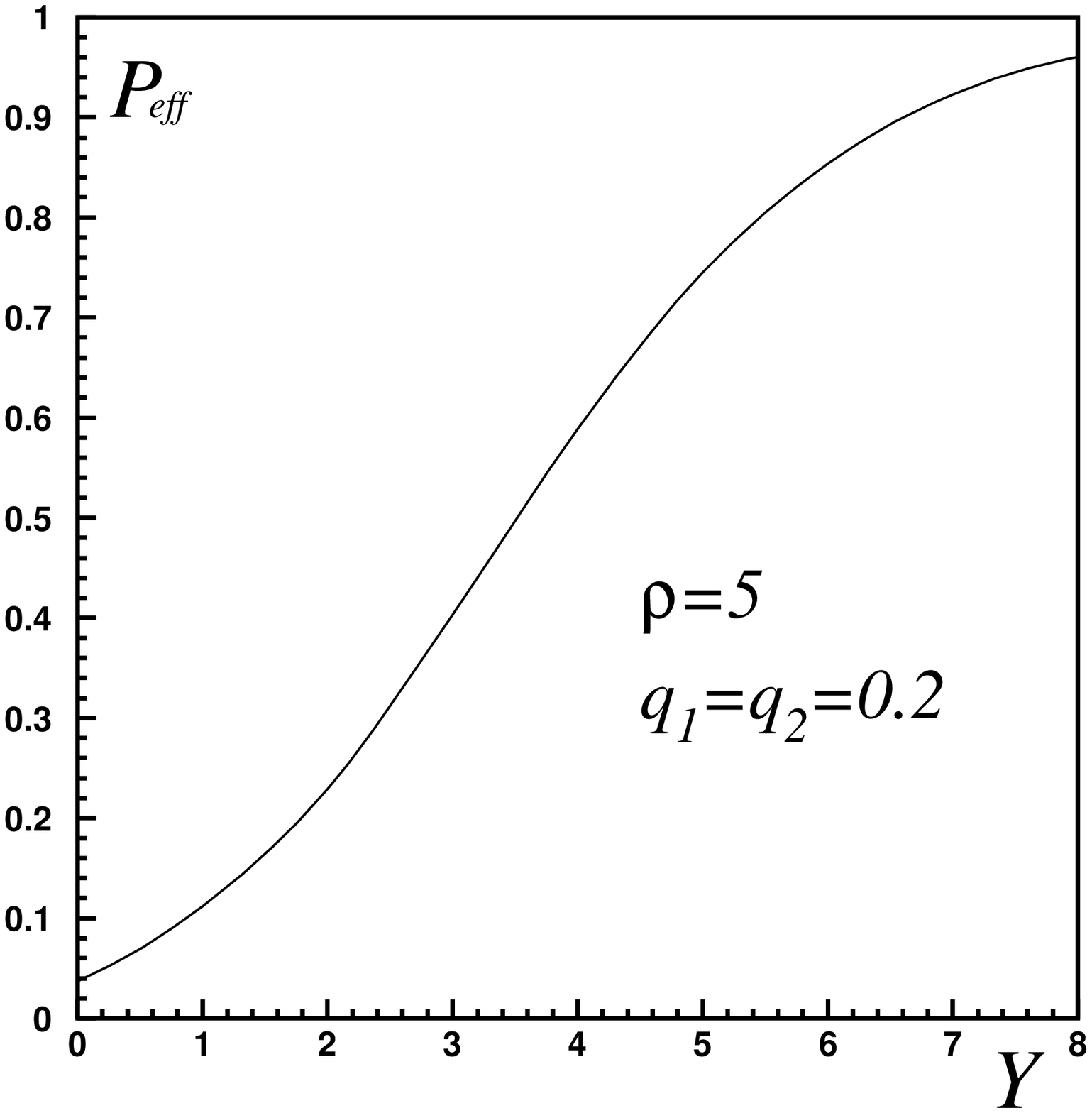,width=85mm}\\
 &  \\ 
\fig{NFV3}-a & \fig{NFV3}-b \\
 &  \\
\end{tabular}
\caption{\it 
The  $P_{amp}(y,q)$ and $P_{eff}(y,q)$ 
functions of the second model
at $\ro\,=\,5\,$  and 
$q_1\,=\,q_{2}\,=1/ \ro\,=0.2\,$ as a functions of scaled
rapidity Y .}
\label{NFV3}
\end{figure}

 The ''effective''  Pomeron propagator we define as 
in the previous model
\beq\label{PFV1} 
P_{eff}(y,q)\,=\,\Le\frac{d\,\Psi(y,q)}{d\,q}\Ra_{q=0}\,\cdot\,q\,\,, 
\eeq
with the initial condition given by
\beq\label{PFV2}
\Psi(y=0,q)\,=\,I(q)\,=\,I(q\,,q_{ext})\,=\,q\,q_{ext}\,.
\eeq
The asymptotic behavior of the $\,P_{eff}(y,q)\,$ at asymptotically large rapidities
is defined by expression Eq.\ref{CVF11}:
\beq
P_{eff}^{asymp.}(q)\,=\,\,\frac{(\ro\,q_{ext})\,(\ro\,q)}
{\,1\,-\,e^{-\ro^2}\,}\,.
\eeq
For the values $q_{ext}\,=\,q\,=\,1\,/\,\ro\,$, we obtain that at large rapidity
this function reaches unitarity limit see plot Fig.\ref{NFV3} and more detailed derivation 
in ~\cite{BMMSX}.
The plot of the Green's function of the theory
\beq\label{Prop222} 
P_{amp}(y,q)\,=\,\frac{P_{eff}(y,q)}{q\,q_{ext}\,},
\eeq
is presented in Fig.\ref{NFV3}.

%%%%%%%%%%%%%%%%%%%%%%%%%%%%%%%%%%%%%%%%%%

\section{Diffractive dissociation process in RFT-0}

\begin{figure}[t]
\begin{center}
\psfig{file=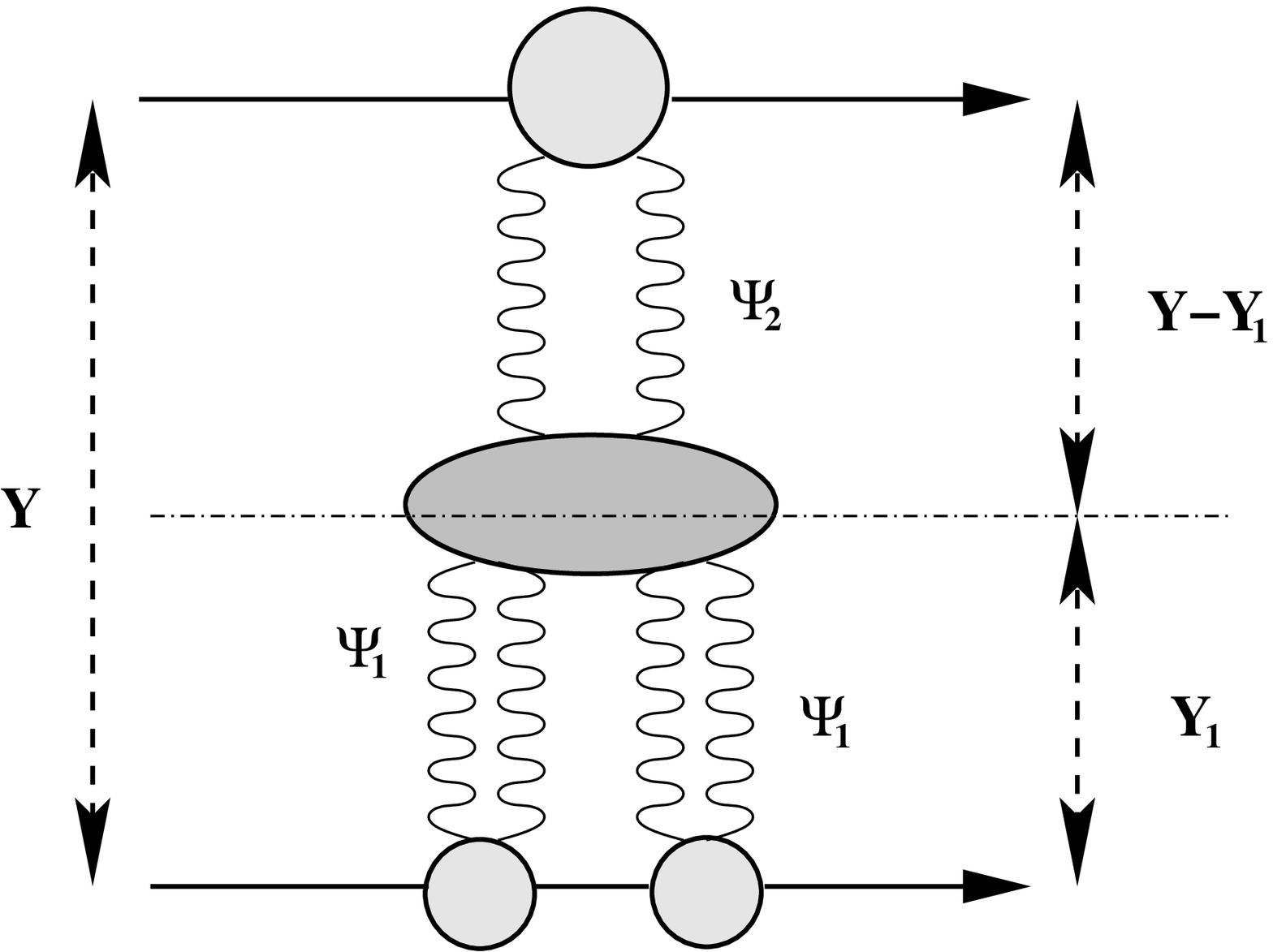,width=110mm} 
\end{center} 
\caption{\it 
The single diffractive dissociation process for rapidity $Y_1$ at total rapidity 
$Y$.}
\label{diffr}
\end{figure}

 As an example of applications of the methods developed
in the paper, we present calculations
of the diffractive dissociation process in the RFT-0 with only 
triple Pomeron vertex.
The algorithm of the calculations is defined as follows.
We have the quantum solution of the theory 
\beq\label{DD1}
\,\Psi_{1}(y,q)\,=\,\sum_{n=0}^{n=\infty}\,\lambda_{n}\,
e^{\,-\,E_{n}\,y}\,\phi_{n}(q)\,\,
\eeq
defined at all range of rapidity $Y$ at some values of the sources
$q_1$ and $q_2$. Let's now consider this solution at the rapidity interval
$Y_1\,<\,Y$. 
With the help of the $\,\Psi_{1}(Y_1,q)\,$
we  define another function $\,\Psi_{2}(y,q)\,$
\beq\label{DD2}
\,\Psi_{2}(y,q)\,=\,\sum_{n=0}^{n=\infty}\,\tilde{\lambda}_{n}\,
e^{\,-\,E_{n}\,\Le y\,-\,Y_1\Ra}\,\phi_{n}(q)\,
\eeq
on the rapidity interval $\,Y_1<\,\,y\,<\,Y\,$ 
with the following initial condition:
\begin{eqnarray}\label{DD3}
\,&\,&\,\Psi_{2}(y=Y_1,q)\,=
\,\sum_{n=0}^{n=\infty}\,\tilde{\lambda}_{n}\,
\,\phi_{n}(q)\,=\,
\,(\,\Psi_{1}(Y_1,q)\,)^{2}\,\,.
\end{eqnarray}
The illustration of this construction see in the 
Fig.\ref{diffr}. 

\begin{figure}[hptb]
\begin{tabular}{ c c}
\psfig{file=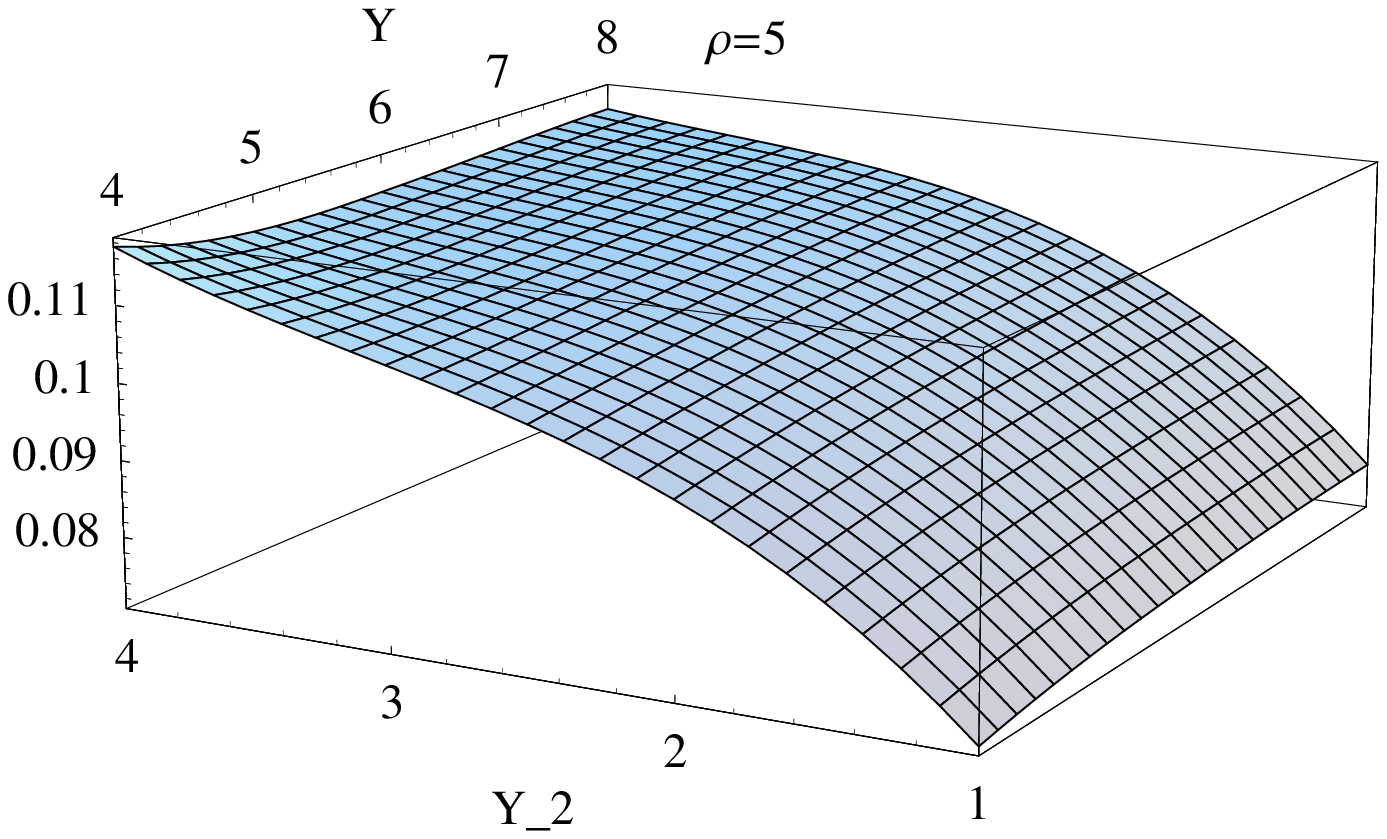 ,width=115mm} & 
\psfig{file=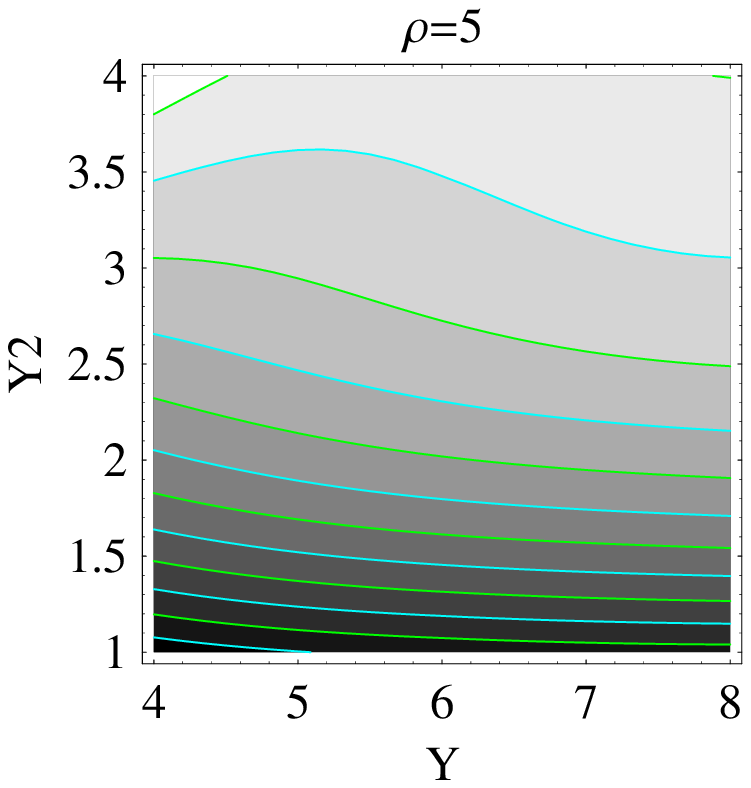,width=60mm}\\
 &  \\ 
\fig{diffr1}-a & \fig{diffr1}-b \\
 &  \\
\psfig{file=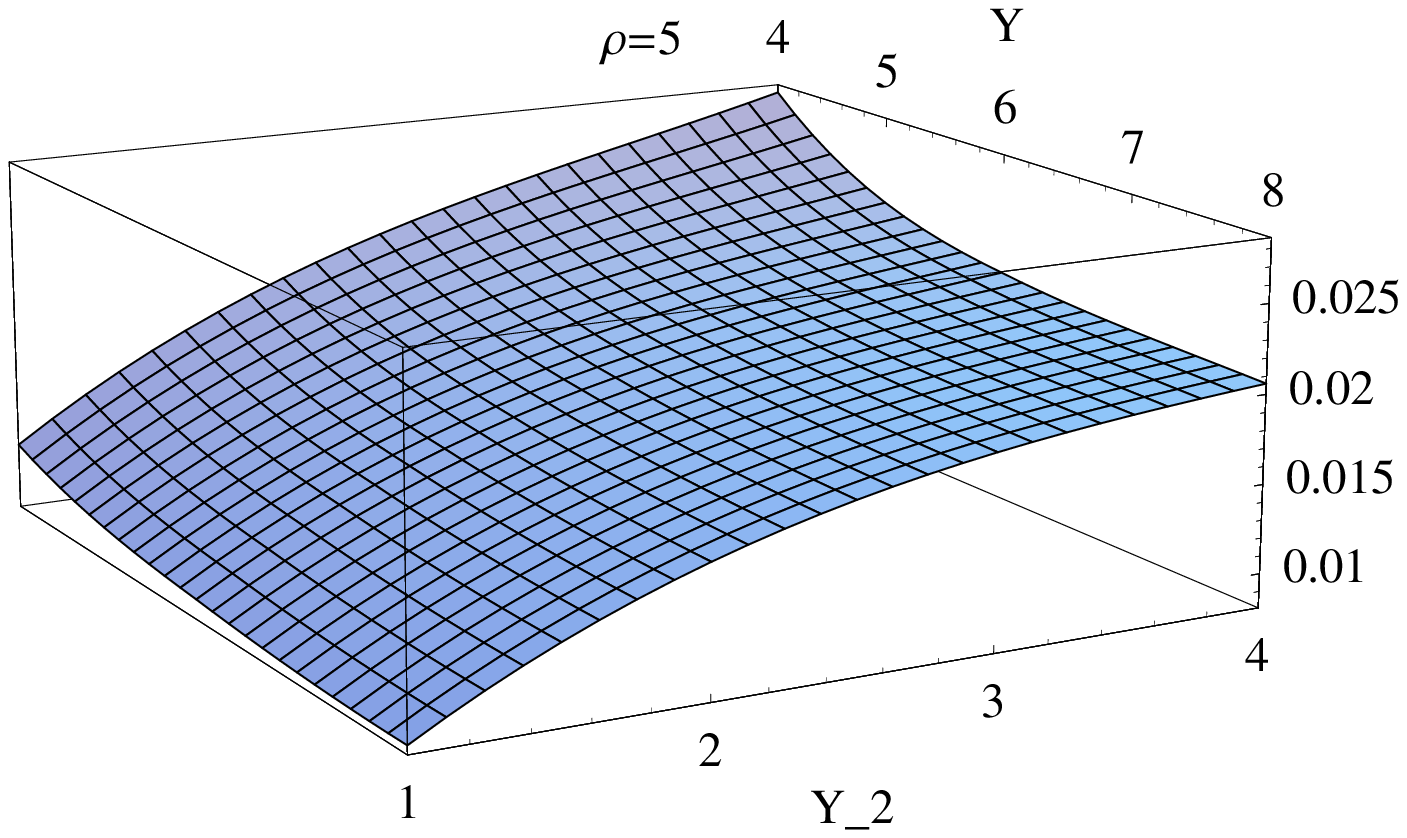 ,width=115mm} & 
\psfig{file=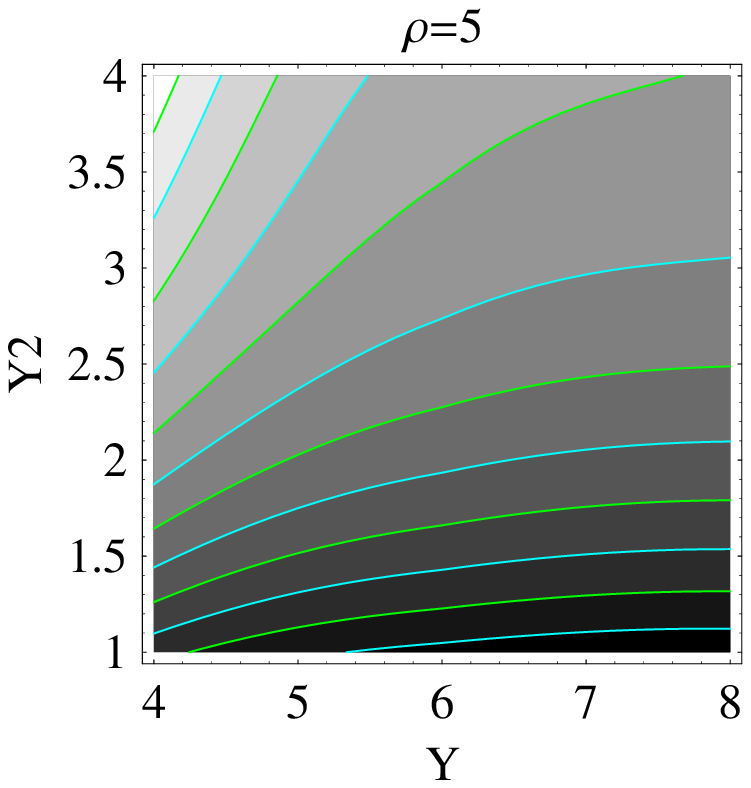,width=60mm}\\
 &  \\ 
\fig{diffr1}-c & \fig{diffr1}-d \\
 &  \\
\end{tabular}
\caption{\it 
The ratio $R$ for the case of symmetrical sources $q_1=q_2=0.2$
in the ~\fig{diffr1}-a - ~\fig{diffr1}-b
and for the symmetrical
sources $q_1=q_2=0.7$ in the 
~\fig{diffr1}-c - ~\fig{diffr1}-d  
correspondingly. The ratios are plotted 
at $\ro\,=\,5\,$  as a functions of scaled
rapidity Y and scaled rapidity variable $Y_2$.}
\label{diffr1}
\end{figure}

\begin{figure}[hptb]
\begin{tabular}{ c c}
\psfig{file=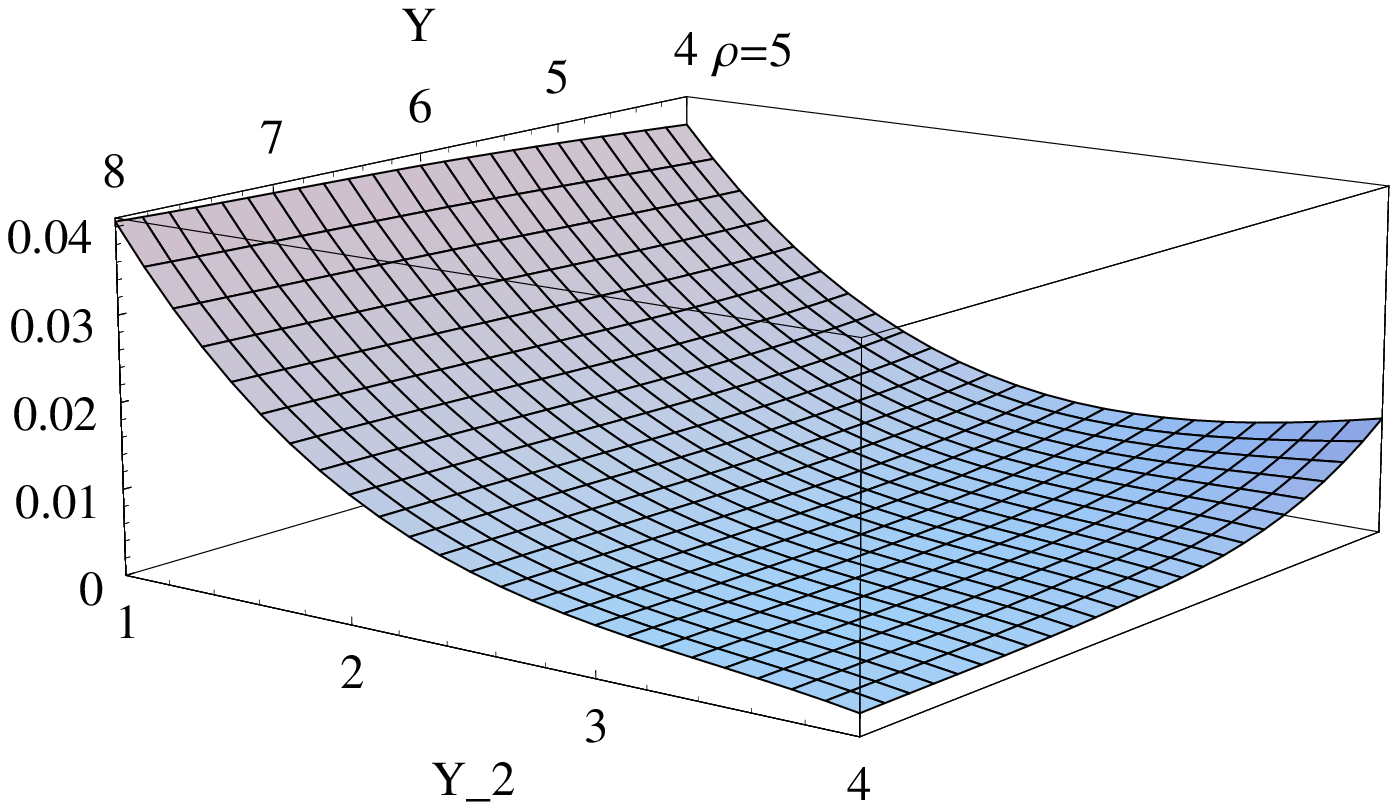 ,width=115mm} & 
\psfig{file=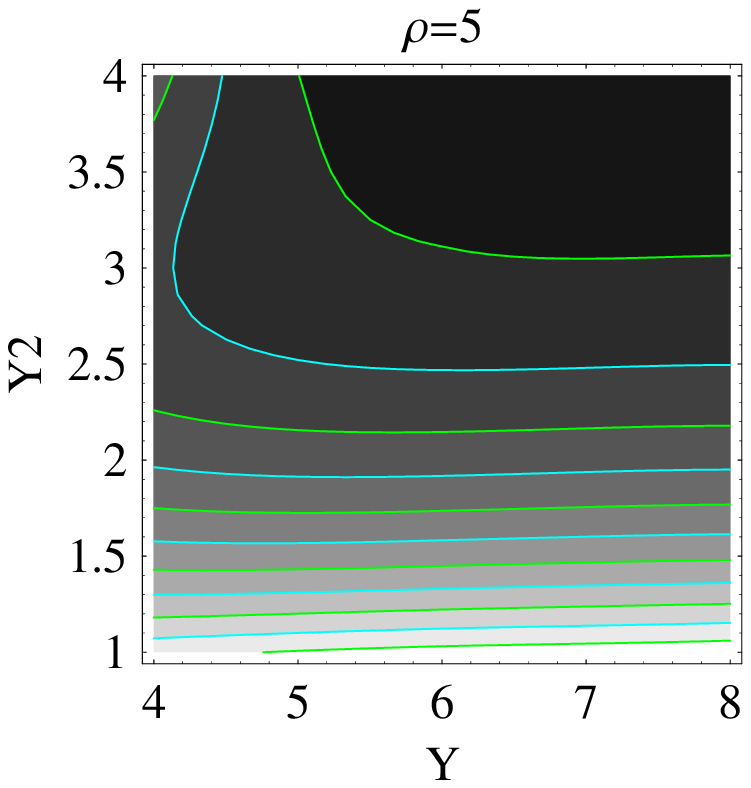,width=60mm}\\
 &  \\ 
\fig{diffr2}-a & \fig{diffr2}-b \\
 &  \\
\psfig{file=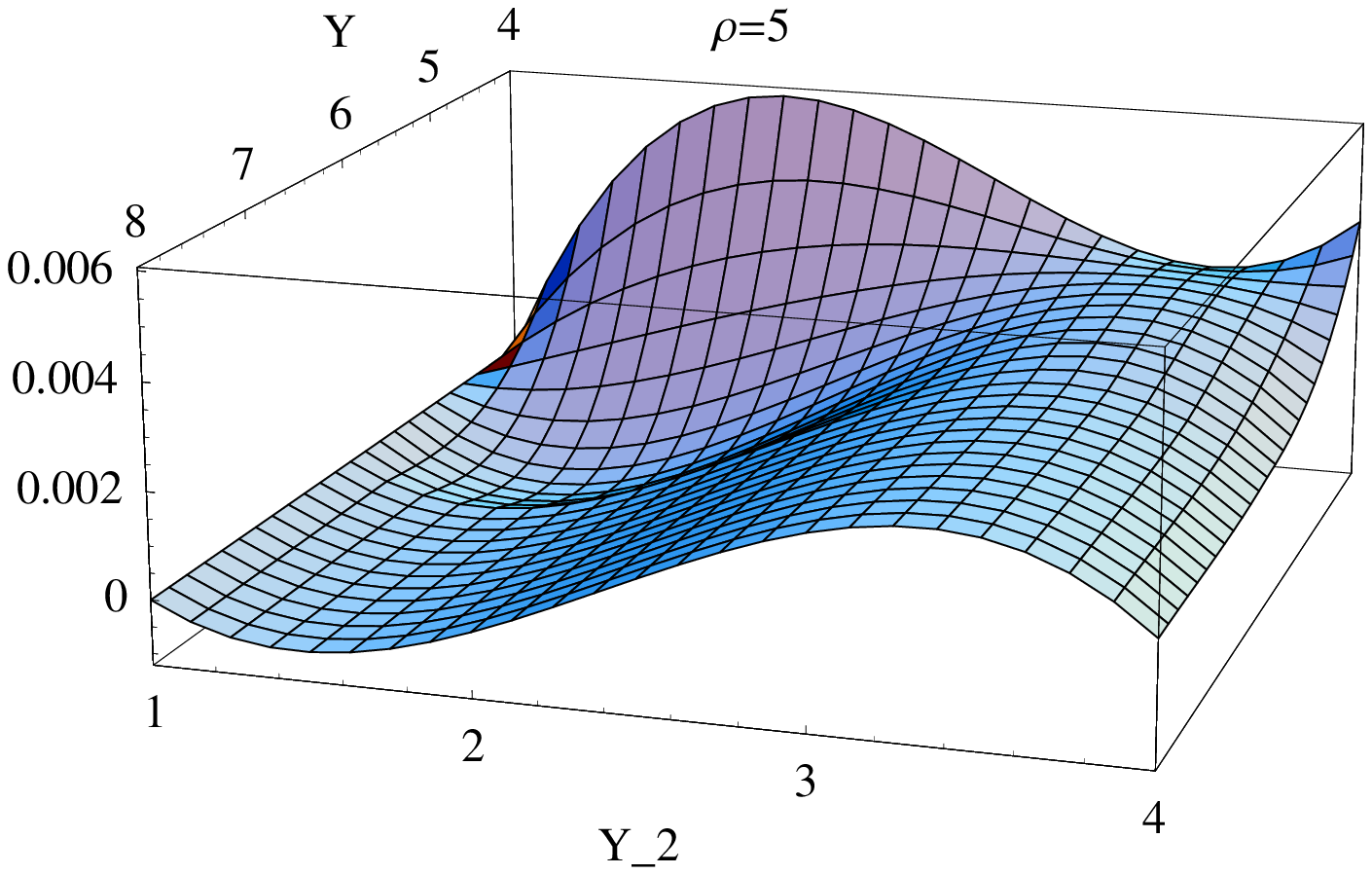 ,width=115mm} & 
\psfig{file=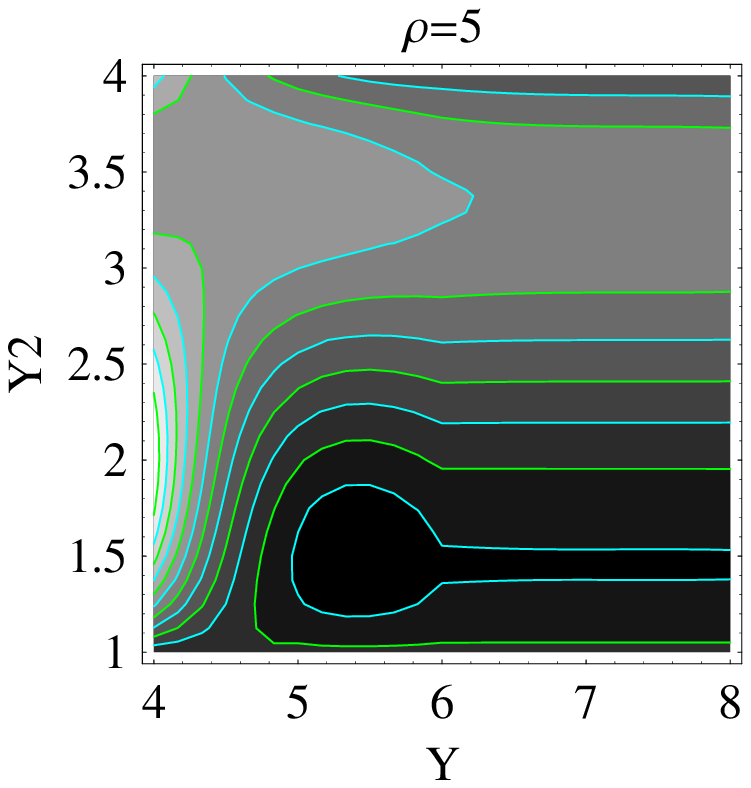,width=60mm}\\
 &  \\ 
\fig{diffr2}-c & \fig{diffr2}-d \\
 &  \\
\end{tabular}
\caption{\it 
The ratio $R_1$ for the case of symmetrical sources $q_1=q_2=0.2$
in the ~\fig{diffr2}-a - ~\fig{diffr2}-b
and for the symmetrical
sources $q_1=q_2=0.7$ in the 
~\fig{diffr2}-c - ~\fig{diffr2}-d 
correspondingly. The ratios are plotted 
at $\ro\,=\,5\,$  as a functions of scaled
rapidity Y and scaled rapidity variable $Y_2$.}
\label{diffr2}
\end{figure}

The only numbers that we need
to calculate now 
are the coefficients $\tilde{\lambda}_{n}\,$. But their calculation
is trivial, they simply
\beq\label{DD4}
\tilde{\lambda}_n\,(Y_1)=\,\frac{\int_{0}^{\infty}\,\phi_{n}(q)\,
(\,\Psi_{1}(Y_1,q)\,)^{2}\,e^{-(q-\ro)^2\,/4}\,
F_{W}(q,\ro)\,dq\,
}{\int_{0}^{\infty}\,\phi_{n}^{2}(q)\,F_{W}(q,\ro)\,
e^{-(q-\ro)^2\,/2}\,dq\,}\,\,,
\eeq
where the weight function $\,F_{W}(q,\ro)\,$
is the same as in the Eq.\ref{Quant9}.
The coefficients $\tilde{\lambda}_{n}\,$ fully determine requested
$\,\Psi_{2}(y,q)\,$ function, which represents the differential cross section of the
single diffraction process 
at given and fixed rapidity interval
$$
y=Y_2=Y-Y_1
$$
Integrating this function over rapidity interval $\,Y_1<\,\,y\,<\,Y\,$
we obtain as the answer the sum of the total diffractive dissociation 
cross sections on this rapidity interval and elastic cross section
\beq
\sigma_{SD}^{Tot}\,+\,\sigma_{el}\,=\,\int_{Y_1}^{Y}\,\Psi_{2}(y,q)\,dy
\eeq
So, the value of the 
single difractive cross section integrated
over rapidity interval $Y_2$ is the following
\beq\label{DD5}
\sigma_{SD}^{Tot}=\,\int_{Y_1}^{Y}\,\Psi_{2}(y,q)\,dy\,-\Le \Psi_{1}(Y,q)\Ra^{2}\,,
\eeq
where $Y$ is the total rapidity of the process,
$Y_1$ is the rapidity "gap" of the process, 
$Y_2$ is the value of the rapidity taken by produced diffractive state and
elastic cross section of the process is defined as
$\sigma_{el}=\Le \Psi_{1}(Y,q)\Ra^{2}$.
In general we do not expect that the numbers given by  RFT-0 theory
will be correct. The only interesting value in the given model, therefore, 
is the ratio of the single diffractive cross section to the 
total cross section.
\beq\label{DD6}
R(Y_1)\,=\,\frac{\sigma_{SD}^{Tot}}{\sigma_{tot}}
\eeq
where $\sigma_{tot}=2\Psi_{1}(Y,q)$.
Considering this ratio we can not prove,
of course, that the same value of the $R$ 
we will obtain in QCD as well, but, nevertheless,  it is interesting to see
what new about this ratio we obtain using full quantum solution for the amplitudes. 
Another interesting ratio, which we can calculate in the given framework, this 
is a ratio of differential single diffraction cross section to the 
total cros section:
\beq\label{DD7}
R_1(Y_1)\,=\,\frac{\Psi_{2}(Y_2)}{\sigma_{tot}}
\eeq
In $R_1$ ratio the dependence on diffractive mass $Y_2$ is present, and it allows
to as to trace the energy dependence of the ratio of the diffractive state 
to the total cross section. Therefore, the following calculations are presented.
We fix the value of diffractive state $Y_2$ equal 1-4 units of 
scaled rapidity as a one variable and as a second variable in the 3D plot
we consider the value of total rapidity $Y$, which we take equal 4-8 units.
These plots we make for the two different symmetrical values of the 
external sources: 0.2 and 0.7.
Obtained plots, see Fig.\ref{diffr1} for the ratio $R$ and
Fig.\ref{diffr2} for the ratio  $R_1$, show the changes of the ration accordingly to
the total rapidity changes.

%%%%%%%%%%%%%%%%%%%%%%%%%%%%%%%%%%%%%%%

\section{Discussion of results}

 The are the following main results of the paper.
First of all, we obtained the 
full quantum solutions for the models and 
compared it with the  classical solutions, illustrating 
the relative contribution of the loops to the amplitude.
In order to find an useful approximation to the full quantum solution
we also considered an eikonalized amplitude which was built with 
the help of the full two point Green's functions. All these amplitudes were calculated
for different values of external vertexes that allows to trace the applicability
of each approximation separately in different parameter's regions.
Another important result is a comparison of quantum solutions with and without quaternary Pomeron vertex
that clarify the contribution of this vertex to the process amplitude .

Plots 
Fig.\ref{EFUN3}-Fig.\ref{EFUN666} represents the result of the calculations in the RFT-0 approach with
triple Pomeron vertex only.
The quantum solution of the model is presented in the plot
Fig.\ref{EFUN3}, whereas the classical solution is presented in the plot Fig.\ref{EFUN4}.
The comparison between these two solutions could be found in the 
plots Fig.\ref{EFUN5}-Fig.\ref{EFUN6}. 
From the Fig.\ref{EFUN3} we see, that our full solution for 
small values of external sources does not reach unitarity really. 
Instead of "black"
disk the "grey" one is achieved. This is may be easily explained
if we will remember that our initial conditions have a eikonal form,
whereas in the s-channel model, see \cite{BMMSX} and references therein, the propagator form of initial
conditions is always applied. In RFT-0 model the black disk limit, as it must be, achieved only at large values of the sources, that
corresponds to the nuclei interaction in usual QCD, see  \cite{braun1,braun2,BGLM}.
Other interesting results are represented in the Fig.\ref{EFUN5}-Fig.\ref{EFUN6}. Indeed, the natural question, which in fact is very 
important in any practical calculations with BK equation involved, is a question about applicability of classical solutions
of the model. Namely, at which values of external sources the difference between the classical and quantum solutions
is acceptable from the point of view of the precision of the amplitude calculations. In real QCD this answer may be obtained only approximately, at least. In RFT-0 calculations the answer is clear, at large values of the external sources the classical solution is acceptable
only at large values of the sources. In the symmetrical case of interactions with the small sources involved, the relative difference between
the amplitudes is very large. The difference pretty fast decreases with the increasing of the value of the sources, but still the difference 
is not negligible 
even when $q=0.3\,-\,0.4$, see Fig.\ref{EFUN5}-Fig.\ref{EFUN6}. Important to underline, that one from the main reasons that provides the applicability 
of the classical solution is a presence of the third classical solution arose at some 
rapidity $Y_c$ ( see Section 2.2 an \eq{Class6}) in the framework. 
The third solution provides the decrease of the classical amplitude
at large values of the total rapidity and without this solution the precision of the model was even worse.  
The situation with the non-symmetrical case of interactions is better, see  Fig.\ref{EFUN66}-Fig.\ref{EFUN666}.
Due the initially broken symmetry of the interaction, the fan diagrams are initially dominant and provide the better
precision of the classical solution in comparison to the symmetrical case of interactions.

Considering the second possible
candidate on the role of "good amplitude approximation", eikonalized function \eq{EikProp}, see Fig.\ref{EFUN10},
we see that at small values of the external sources this amplitude is indeed better than the classical solution. In the region
of large values of the sources both amplitude are equal more or less - the unitarization corrections are small 
for the large sources. We also could compare the effective Pomeron propagator 
with the asymptotic results obtained  in the framework of s-channel model, see \cite{BMMSX}.
Looking in the 
Fig.\ref{EFUN8} and Fig.\ref{NFV3}, we see, that the value of the 
full propagator, which considered as an amplitude in s-channel model, is close to one, especially
for the second considered model with quaternary vertex included, which has direct relations with
s-channel model,  see again \cite{BMMSX}. 
But, in general, for an arbitrary value of a source there is no
propagator's unitarization and the way to achieve the  unitarization in this
case is the eikonalization of the propagator only.

Interesting problem, which also could be investigated in the framework of RFT-0, is a problem of the
influence of the value of the triple Pomeron vertex on the behavior of the amplitude. Changing value
of the $\rho$ parameter from the $5$ till $1$ we change the value of the triple vertex $\lambda$
from $0.04$ till $0.1$. The resulting amplitude is depicted in the Fig.\ref{EFUN7}. From this plot we see, 
that already at relatively small values of rapidity this amplitude is decreasing till zero. 
This fast of the decreasing of the amplitude at
large values of $\lambda$ denote a main difference between the models of RFT-0 with and without 
quaternary vertex. In RFT-0 without the quaternary vertex the amplitude, as it mentioned above, approaches zero
whereas in RFT-0 with the quaternary vertex the amplitude approach constant with any value of
triple vertex and correctly adjusted value of quaternary vertex, see \eq{FV33}, \eq{CVF9} and \eq{CVF11}.
Of course, when the value of the triple vertex is small, two models do not so differ, see
Fig.\ref{NFV2}. This difference in the behavior of the amplitudes in two models may be 
also analyzed in the terms of s-channel model, see \cite{BMMSX} for the details and explanations.

As a possible example of the application of the model we calculated 
the values of the ratios of the single diffractive and differential single diffractive cross
sections to the total cross section in the given framework, see Eq.\ref{DD6} and Eq.\ref{DD7}.
The results of these calculations are present in the
Fig.\ref{diffr1} - Fig.\ref{diffr2}. The plot
Fig.\ref{diffr1} represents the ratio of single diffractive cross section
(integrated over rapidity of the diffractive state) to the value of the total cross
section. In two cases when the sources are equal 0.2 and 0.7, we obtain, that
for the fixed value of the diffractive state this ratio almost does not change with energy.
More precisely,  in both cases there is a tiny change in the ratio behavior
at small rapidities 4-5, and constant behavior at large rapidiies 6-8.
Tracing the relative contribution of the diffractive state at fixed rapidity
in both cases we see, that the contribution of the
large mass diffractive state, $Y_2=4$, approximately
two times large than the contribution from small , "elastic", diffractive state
$Y_2=1$,
as it must be in reality. We see, that the simple RFT-0 model correctly reproduce
main futures of the real QCD. It is also interesting to note, that
the relative contribution of the large mass diffractive state is larger for the case of
small external sources, there is $R=0.11$  for source $q=0.2$ against
$R=0.028$  for source $q=0.7$. 
Concerning the calculations of the second ratio,  $R_1$, see Fig.\ref{diffr2},
we obtain that these two cases are different. At small source, $q=0.2$, the maximum of the ratio is at
small diffractive state, which has almost constant behavior with total rapidity.
At large vale of source, $q=0.7$ , the situation is opposite, the maximim contribution
comes from the large mass diffractive state. The contribution of the small diffractive state
at $q=0.7$ is zero in Fig.\ref{diffr2} at large values of Y, that means 
constant value of single diffraction cross section at small values of $Y_2$
and large values of $Y$. In general, considering Fig.\ref{diffr1} - Fig.\ref{diffr2},
we can conclude, that the description of the diffractive states
in RFT in the case when we include in calculations all possible re-scattering corrections
is different from the "naive" diffraction models, where only part of corrections
is included. Therefore, there is a hope, that using the same
receipts of calculations in the QCD RFT, we will be able correctly describe diffraction
data and their energy dependence. 

%%%%%%%%%%%%%%%%%%%%%%%%%%%%%%%%%%%%%%%%%%

\section{Conclusion}

In this paper we performed calculations of the different amplitudes in the framework of the RFT-0 model.
Comparing different approximation we analyze the applicability of the different
approximation schemes and their dependence on the parameters of the model. The main conclusion of the calculations is that we could trust 
to the classical approximation for the amplitude only when the value of the triple Pomeron vertex is small.
In this region the difference between models with and without quaternary vertexes is negligible. Nevertheless, the important condition for this
approximation in the case of interacting of symmetrical particles is an including of the third classical solution in the set of classical solutions, which arises at some critical value 
of rapidity. Without this solution the classical approximation is not good, even at not small values of the external sources.
For the case of the small values of these vertexes the eikonalized Green's function amplitude is more precise solution then the classical one. Unfortunately,
in real QCD this amplitude also could not be calculated precisely and therefore we could not trust to the classical solutions at all when the
values of the external sources is not large enough. Does the proton is "large enough" in the case of real QCD is an open question,
which could not be answered, unfortunately, in the given framework. 

When the value of the triple vertex is growing, 
the picture is changing drastically.
In this case we could not trust anymore to the classical solutions of the models. Also, there is
a large difference in the behavior of the amplitude in models with and without quaternary vertex, it's presence
changes the asymptotic behavior of the amplitude. 
In real QCD it could means, that the different evolution equations must be
applied in the different regions of the impact parameter space with the different values of the coupling constant. 
If we assume, that the influence of the non-perturbative effects is in the change of the value of the
coupling constant only, then we must separate the contribution of the perturbative spots in impact parameter space,
which evolution is described by the BK equation, from the non-perturbative regions where different evolution equations must be applied.
In this case the overall amplitude is a sum of the amplitudes described by the different evolution equations in different regions, 
with the BK equation applicable in the regions of high partons density only.
This picture will lead to the factorization of the non-perturbative effects from the perturbative ones, first of all, and also it
could explain the possible applicability of the BK equation in case of proton-proton collisions. Namely, at high enough energies
the overall contribution of "black" spots may be larger then the contribution of the "white" and "grey" ones and 
we could describe even inclusive data by the BK equation formalism. The large but constant non-perturbative contributions
into the amplitude in this case may be accounted by the adjusting of the values of the external sources in interaction of interests.

Concluding we underline, that RFT-0 model is a very interesting and important ground for an initial implementation 
of the different ideas which further could be applied in real QCD calculations as well.
 
%%%%%%%%%%%%%%%%%%%%%%%%%%%%%%%%%%%%%%%%%%

\section*{Acknowledgments}

I am grateful to Leszek Motyka for his participation in the development
of the main themes of the paper and to M.Braun for his support and interest to the paper during
period of it's writing.

%%%%%%%%%%%%%%%%%%%%%%%%%%%%%%%%%%%%%%%%%%

\end{document}